\documentclass[11pt,a4paper]{article}
\usepackage[dvipdfmx]{graphicx}
\usepackage{amsmath, amssymb}
\usepackage{color}
\usepackage{ulem}

\newcommand{\vct}[1]{\boldsymbol{#1}}
\newcommand{\eden}{\mbox{\Large $\varepsilon$}}
\newcommand{\eb}{E_{\rm B}}

\begin{document}

\baselineskip=16pt

{
\noindent
\Large\textbf{%
The relationship of the neutron-skin thickness\\[4pt]
to the symmetry energy and its slope
}
}

\vspace{\baselineskip}

\noindent
Toshio Suzuki\footnote{kt.suzuki2th@gmail.com}

\vspace{0.5\baselineskip}

\noindent
\parbox[t]{14cm}{
Research Center for Electron Photon Science, Tohoku University,\\[2pt]
Sendai 982-0826, Japan
}

\begin{center}
\parbox[t]{14cm}{
\small

\baselineskip=12pt

The neutron-skin thickness of asymmetric semi-infinite nuclear matter($\delta R_M$)
is shown to be a function of
Coulomb energy($V_c$), the asymmetry-energy coefficient($J$),
the slope($L$) of the asymmetry energy, and the incompressibility
coefficient($K$),
in addition to he Fermi momentum($k_{\rm F}$) and the asymmetry parameter($I=(N-Z)/A$). 
The relational formula is derived on the basis of the Hugenholtz-Van Hove theorem
in the mean-field(MF) approximation for nuclear matter.
Using the formula as a guide, the neutron-skin thickness($\delta R$)
in $^{208}$Pb is examined in the MF models.
The linear correlation between $L$ and $\delta R$ appears as a kind of
spurious ones through the model-dependent correlation
of $L$ with $J$ which is included in the main components of the formula.
}
 
\end{center}

\vspace{0.5\baselineskip}

\section{Introduction}\label{intro}

The neutron distribution is one of the most fundamental quantities to 
understand nuclei, together with the proton distribution\cite{bm}.
Hence, the neutron-skin thickness($\delta R=R_n-R_p$) has been discussed for a long time
in nuclear physics\cite{my,br}, where $R_\tau$ represents the root msr(mean square radius)
of the point neutron($\tau=n$) and proton($\tau=p$) distributions, respectively. 
So far, however, the arguments on the neutron-skin thickness(NST)
have not been qualitative enough to understand its structure\cite{br},
because of the difficulty to observe  the neutron distribution experimentally\cite{thi}.
The neutron distribution has been explored mainly through hadron interaction, where
the interaction and reaction mechanism are not defined without ambiguity\cite{thi},
while the proton distribution through the electromagnetic one which is well understood
theoretically\cite{def,bd,ks1}.

Recently, interest in the NST($\delta R$) is increased
experimentally\cite{thi,jlab1,jlab2}.
and theoretically\cite{rein0,roca1,roca2,roca3,kss,ks2}.
On the one hand, aiming to observe the neutron distribution,
the parity-violating electron scattering, which is long awaited\cite{sick},
has been performed\cite{jlab1,jlab2},
where the interaction and reaction mechanism are well understood
as in the conventional electron scattering.
On the other hand, it has been pointed out numerically that
the values of $\delta R$ are correlated with those of
the slope ($L$) of the asymmetry energy($J(\rho)$) as a function of
the nucleon density($\rho$)\cite{rein0}.
Refs.\cite{roca1,roca2,roca3} have
calculated the values of $L$ and  $\delta R$ in $^{208}$Pb using a large
number of nuclear models, and plotted the point $(L, \delta R)$ obtained in each model
in the $(L-\delta R)$-plane. By analyzing those points with the least squares analysis(LSA),
it has been shown that almost all the points are well on the least square line(LSL).
This fact implies a possibility to determine the value of $L$ by the observed value
of $\delta R$\cite{thi,jlab1,jlab2,roca1, roca2,roca3}.

Unfortunately, however, the above LSL is mode-dependent. Other calculations have
provided different LSL's\cite{na,jlab3}.
Moreover, it is not clear yet
why $\delta R$ calculated
in the various nuclear models is proportional to $L$.
The proportionality is shown numerically
and is interpreted qualitatively on the basis of the droplet model only\cite{roca1}.

The purpose of the present paper is two-fold. The one is to derive the relational
formula of $\delta R$ with macroscopic quantities
of infinite nuclear matter like $L$, without invoking liquid-drop assumptions.
The other is to investigate the correlations of $\delta R$ calculated
in the mean field(MF) models with the macroscopic quantities
by the LSA on the basis of the obtained relational formula.  
For this purpose, the present paper is organized as follows.
In the next section, the asymmetric semi-infinite nuclear matter(SINM)
used in Refs.\cite{br,ks2} will be briefly reviewed.
Then, the NST of the SINM($\delta R_M$) will be shown to be expressed 
in an analytic way as a function of $L$, the asymmetry energy coefficient($J$),
the incompressibility coefficient($K$) and the Coulomb energy($V_c$),
in addition to the Fermi momentum($k_{\rm F}$) and the asymmetry parameter($I=(N-Z)/A$).
The relational formula of $\delta R_M$, which will be called the NST formula,
is derived from the Hugenholtz-Van Hove(HVH)
theorem in the MF approximation\cite{bethe,weisskopf,hvh}.
In \S 3, the NST of finite nuclei$(\delta R)$ will be discussed,
on the basis of the NST formula. First, $\delta R$ for the Fermi-type distributions
of neutrons and protons is separated into two parts
as $\delta R=\delta R_0+\delta R_a$.
The term $\delta R_0$ will be shown to correspond to $\delta R_M$,
and $\delta R_a$ is due to the diffuseness part of the Fermi-type function.
Second, using the results in Ref.\cite{ks2} where the neutron and proton distributions
obtained in the MF models are approximated by the Fermi-type function,
$\delta R_0$ in $^{208}$Pb is analyzed according to the NST formula.
It will be seen that the contribution of $L$ and $K$ to $\delta R_0$
is less than $10\%$, while that from $J$ together with $V_c$ and $I$ dominates $\delta R$.
Third, the correlations between $\delta R$  and the macroscopic quantities in the MF models
will be explored by LSA. It will be shown that the linear relationship
between $J$ and $\delta R$ is obtained, whereas the one between $L$ and $\delta R$
appears as a kind of spurious correlations through the correlation of $L$ with $J$
in the MF models.
As a result, even if the value of $\delta R$ is
determined experimentally, the value of $L$ remains to be a free-parameter
in the MF models. 
The final section will be devoted to a brief summary.

\section{Asymmetric semi-infinite nuclear matter and the mean square radius}\label{ma}

\subsection{Asymmetric semi-infinite nuclear matter}\label{anm}

The symmetric nuclear matter is widely used as a guide in investigating the structure
of finite nuclei\cite{bm}.
As far as the author knows, the asymmetric nuclear matter has been
assumed rarely in nuclear
physics, since it is necessary to take into account of the Coulomb energy.
Recently, Ref.\cite{ks2} has employed a simple model to describe asymmetric
semi-infinite nuclear matter(SINM) with the Coulomb energy\cite{br}.
It has been shown that the SINM model is useful to understand in a systematic way
the results calculated in the MF models 
for the stable finite nucleus like $^{208}$Pb\,\cite{ks2}.

The SINM model assumes the total nuclear energy density to be described as\cite{br}
\begin{equation}
\eden_{\rm asym}=\eden+V_c\rho_p, \label{med}
\end{equation}
where $V_c\rho_p$ stands for the ``Coulomb term'', $\rho_p$ being the proton density
and $V_c$ a constant.
The energy density $\eden$ is provided for nuclear matter, for example,
by the relativistic MF(RMF) or non-relativistic Skyrme-type MF(SMF) models\cite{ks2}.
The value of the constant $V_c$ is approximately given by the value at the
center in the Coulomb potential of the uniformly charged sphere
with the radius of $r_cA^{1/3}$(fm)\cite{br,ks2},
\begin{equation}
 V_c=\frac{3}{2}\frac{Z\alpha}{r_cA^{1/3}}.\label{coulomb}
\end{equation}
Ref.\cite{ks2} has used $V_c=22.144$ MeV for $^{208}$Pb with $r_c=1.350$ fm.

The Hugenholtz and Van Hove(HVH) theorem\cite{bethe,weisskopf,hvh}
for symmetric matter is extended to asymmetric matter by requiring  
\begin{equation}
\frac{\partial}{\partial\rho_\tau}\left(\frac{\eden_{\rm asym}}{\rho}\right)=0.\label{me}
\end{equation}
This yields the relationship between the binding energy per nucleon($\eb$)
and Fermi energy($E_{{\rm F}_\tau}$) of neutrons and protons, as in HVH theorem
for symmetric matter, but with $V_c$ as
\begin{equation}
 \eb
 =E_{{\rm F}_\tau}
 =\left\{
 \begin{array}{ll}
E_{{\rm F}_{0,n}}, &\tau=n\\[2pt]
E_{{\rm F}_{0,p}}+V_c,\,&\tau=p
\end{array}
\right.
\quad , \quad E_{{\rm F}_{0,\tau}}= \frac{\partial\eden}{\partial\rho_\tau}.\label{mhvh1}
\end{equation}
The above equation is given for non-relativistic models, while for relativistic models,
$E_{{\rm F}_\tau}$ is replaced by ($E_{{\rm F}_\tau}-M$), $M$ being the rest mass
of the nucleon.
Eq.(\ref{mhvh1}) ensures that the SINM model is suitable for being used
as a guide in discussing a stable finite nucleus with the Coulomb potential 
where protons and neutrons have the same Fermi energy.

\subsection{Mean square radius and neutron-skin thickness}

The neutron and proton distributions in SINM are assumed with the uniform-density sphere
with nucleon number $A=N+Z$ as 
\begin{equation}
 A=\frac{4\pi}{3}\rho_0R_0^3, \qquad N_\tau=\frac{4\pi}{3}\rho_\tau R^3_{0,\tau}\,,
\end{equation}
where $\rho_0$ denotes the nucleon density given in $V_c=0$, while $\rho_\tau$
in $V_c\ne 0$.
The $N_\tau$ represents the neutron number $N$ for $\tau=n$,
and the proton number $Z$ for $\tau=p$.
Then root msr is given by
\begin{equation}
 R_\tau=\sqrt{\frac{3}{5}}R_{0,\tau}=\sqrt{\frac{3}{5}}
 \left(\frac{3N_\tau}{4\pi\rho_\tau}\right)^{1/3},\label{r}
\end{equation}
In expanding $N_\tau$ by the asymmetry factor $I=(N-Z)/A$ as
\begin{equation}
 \frac{N}{A}=\frac{1}{2}\left(1+I\right), \quad
 \frac{Z}{A}=\frac{1}{2}\left(1-I\right),
\end{equation}
$\delta R_M$ of SINM is written as 
\begin{equation}
\delta R_M =\sqrt{3/5} \left(\frac{3A}{8\pi}\right)^{1/3}\left((\frac{1}{\rho_n^{1/3}}
							-\frac{1}{\rho_p^{1/3}})
+\frac{I}{3}(\frac{1}{\rho_n^{1/3}}+\frac{1}{\rho_p^{1/3}})\right).\label{delta}
\end{equation}
The density $\rho_\tau$ is expanded in terms of $V_c$ up to order of $O(V_c^2)$ as
\begin{equation}
\rho_\tau\approx\rho_{0,\tau}(1+\alpha_\tau V_c+\frac{1}{2}\beta_\tau V_c^2),\quad
(\, \rho_{0,\tau}=\frac{1}{2}\rho_0 \,).\label{dex}
\end{equation}
The coefficients, $\alpha_\tau$ and $\beta_\tau$, are determined through 
Eq.(\ref{me}) from the HVH theorem as follows.
First, $\eden_{\rm asym}/\rho (\rho=\rho_n+\rho_p)$ is expanded as 
\begin{equation}
\frac{\eden_{\rm asym}}{\rho}=\frac{\eden}{\rho_0}
 +\frac{1}{2}V_c+\frac{1}{4}\left(-(\alpha_p+\alpha_n)
+\frac{1}{2}\rho_0(\alpha^2_pc_{pp}+2\alpha_p\alpha_nc_{pn}+\alpha^2_nc_{nn})
+2\alpha_p\right)V_c^2.\label{e1}
\end{equation}
Here, $c_{\tau.\tau'}$ is defined by
\begin{equation}
c_{\tau,\tau'}=\frac{\partial^2 \eden}
{\partial \rho_\tau\partial\rho_{\tau'}}|_{\rho=\rho_0,\rho_3=0}\,, \quad \rho_3=\rho_n-\rho_p,
\end{equation}
which is rewritten in terms of $J$ and $K$  as
\begin{equation}
c_{pp}=c_{nn}=\frac{1}{\rho_0}(\frac{K}{9}+2J),\quad
 c_{pn}=c_{np}=\frac{1}{\rho_0}(\frac{K}{9}-2J), \label{jk}
\end{equation}
by their definitions,
\begin{equation}
J=\frac{\rho}{2}\frac{\partial^2 \eden}{\partial\rho^2_3}|_{\rho=\rho_0,\rho_3=0}\,,\quad
K=9\rho\frac{\partial^2\eden}{\partial \rho^2}|_{\rho=\rho=0}\,.
\end{equation}
Next, $\partial \eden_{\rm asym}/\partial \rho_\tau$ is expanded as
\begin{align}
\frac{\partial \eden_{\rm sym}}{\partial \rho_p}
 &=\frac{\eden}{\rho_0}+\left(1+\frac{\rho_0}{2}
	(\alpha_pc_{pp}+\alpha_nc_{np})\right)V_c \nonumber\\
 &+\frac{1}{4}\left(\rho_0(\beta_pc_{pp}+\beta_nc_{np})
+\frac{\rho^2_0}{2}
	     \frac{\partial}{\partial \rho_p}
(\alpha^2_p\frac{\partial^2}{\partial\rho^2_p}+2\alpha_p\alpha_n\frac{\partial^2}
{\partial \rho_n\partial\rho_p}+\alpha_n^2\frac{\partial^2}{\partial\rho^2_n})
 \eden |_{\rho=\rho_0,\rho_3=0}    \label{e2}
	       \right)V^2_c,\\
 \frac{\partial \varepsilon_{\rm sym}}{\partial \rho_n}
 &=\frac{\eden}{\rho_0}+\frac{\rho_0}{2}
	(\alpha_nc_{nn}+\alpha_pc_{pn})V_c \nonumber\\
 &+\frac{1}{4}\left(\rho_0(\beta_nc_{nn}+\beta_pc_{pn})+\frac{\rho^2_0}{2}
	     \frac{\partial}{\partial \rho_n}
(\alpha^2_p\frac{\partial^2}{\partial\rho^2_p}+2\alpha_p\alpha_n\frac{\partial^2}
{\partial \rho_n\partial\rho_p}+\alpha_n^2\frac{\partial^2}{\partial\rho^2_n})
 \eden |_{\rho=\rho_0,\rho_3=0}\label{e3}
	       \right)V^2_c.
\end{align}
By requiring Eq.(\ref{me}) which yields
\begin{equation}
\frac{\varepsilon_{\rm sym}}{\rho}=\frac{\partial \varepsilon_{\rm sym}}{\partial \rho_p}
 = \frac{\partial \varepsilon_{\rm sym}}{\partial \rho_n},
\end{equation}
the first-order terms of $V_c$ in Eqs.(\ref{e1}), (\ref{e2}) and (\ref{e3}) provide the equations,
\begin{equation}
\rho_0(\alpha_pc_{pp}+\alpha_nc_{np})=-1, \quad
\rho_0(\alpha_nc_{nn}+\alpha_pc_{pn})=1 , \label{v1}
\end{equation}
which determine $\alpha_\tau$, together with Eq.(\ref{jk}), as
\begin{equation}
\alpha_p=-\alpha_n=-\frac{1}{4J}.\label{al}
\end{equation}
The second order terms of $V_c$ in Eqs.(\ref{e1}), (\ref{e2}) and (\ref{e3})  give
\begin{align}
\rho_0\alpha^2_p(c_{pp}-c_{pn})+2\alpha_p
&=\rho_0(\beta_pc_{pp}+\beta_nc_{np})+\frac{\rho^2_0}{2}
	  \alpha^2_p\frac{\partial}{\partial \rho_p}
(\frac{\partial^2}{\partial\rho^2_p}-2\frac{\partial^2}
{\partial \rho_n\partial\rho_p}+\frac{\partial^2}{\partial\rho^2_n})\eden|_{\rho=\rho_0,\rho_3=0},
\label{v211}\\
 \rho_0\alpha^2_p(c_{pp}-c_{pn})+2\alpha_p
&=\rho_0(\beta_nc_{nn}+\beta_pc_{pn})+\frac{\rho^2_0}{2}
	  \alpha^2_p\frac{\partial}{\partial \rho_n}
(\frac{\partial^2}{\partial\rho^2_p}-2\frac{\partial^2}
{\partial \rho_n\partial\rho_p}+\frac{\partial^2}{\partial\rho^2_n})\eden|_{\rho=\rho_0,\rho_3=0},
\label{v222}
\end{align}
where $\alpha_n+\alpha_p=0$ has been used, according to Eq.(\ref{al}).
The above two equations yield
\begin{align}
2\rho_0\alpha^2_p(c_{pp}-c_{pn})+4\alpha_p
&=\rho_0(\beta_p+\beta_n)(c_{pp}+c_{np})
+4\rho^2_0\alpha^2_p\frac{\partial}{\partial \rho}
\frac{\partial^2}{\partial\rho^2_3}\eden|_{\rho=\rho_0,\rho_3=0},\label{p}\\
0&=\rho_0(\beta_p-\beta_n)(c_{pp}-c_{np})
+4\rho^2_0\alpha^2_p\frac{\partial}{\partial \rho_3}
\frac{\partial^2}{\partial\rho^2_3}\eden|_{\rho=\rho_0,\rho_3=0}.\label{m}
\end{align}
Eqs.(\ref{jk}) and (\ref{p}) provide
\begin{equation}
\frac{\rho_0}{4}(\beta_p+\beta_n)=
-\frac{3\rho_0}{16J^2K}L,\quad L=3\rho\frac{\partial}{\partial \rho}\frac{\rho}{2}
\frac{\partial^2\eden}{\partial\rho^2_3}|_{\rho=\rho_0,\rho_3=0},\label{l}
\end{equation}
while Eq.(\ref{m}) gives
\begin{equation}
\frac{1}{4}\rho_0(\beta_p-\beta_n)= 
-\rho^3_0\frac{1}{64J^3}\frac{\partial}{\partial \rho_3}
\frac{\partial^2}{\partial\rho^2_3}\eden|_{\rho=\rho_0,\rho_3=0}=0,\label{mm}
\end{equation}
yielding $ \beta_p=\beta_n$.
Consequently, we obtain
\begin{equation}
\beta_\tau=-\frac{3}{8J^2K}L.\label{be}
\end{equation}

Finally, Eqs.(\ref{delta}) and (\ref{dex}) with Eqs.(\ref{al}) and (\ref{be})
provide us with the expression of the NST of SINM as
\begin{equation}
\delta R_M=\sqrt{\frac{3}{5}}\frac{2}{3}\left(\frac{9\pi A}{8}\right)^{1/3}\frac{1}{k_{\rm F}}
\left(I-\frac{1}{4}\left(\frac{V_c}{J}\right)+\frac{I}{24}\left(\frac{3L}{2K}
		  +\frac{1}{3}\right)\left(\frac{V_c}{J}\right)^2\right),\label{result}
\end{equation} 
where $\rho_0$ is replaced by $(2k_{\rm F}^3/3\pi^2)$, $k_{\rm F}$ being the Fermi
momentum corresponding to $\rho_0$, since most of the MF models employ the value
of $k_{\rm F}$ in order to fix the values of their parameters
as one of the inputs\cite{nl3,sly4}.
Up to the first-order term of $V_c$, the above result is similar to the one of the droplet model
neglecting the surface effect\cite{my,br}.
Eq.(\ref{result}) is named the NST formula.

Before detailed numerical investigations of Eq.(\ref{result}), it may be worthwhile making
several qualitative comments. 
First, the NST formula depends on several nuclear matter parameters
like $J$, $L$ and $K$. This fact implies that even if the experimental value of NST is
obtained, the values of macroscopic parameters are not determined model-independently. 
Second, the proportional relation of $\delta R_M$ to $I$ is expected naturally, and indeed
has been observed in the numerical calculations based on the MF models\cite{roca2}.
Third, if the value of $V_c$ is taken to be equal to zero, Eq.(\ref{mhvh1})
from the HVH theorem provides the same Fermi momentum for neutrons and protons as
$k_{{\rm F}_n}=k_{{\rm F}_p}$. As a result, one has $N=Z$ and $\delta R_M =0$.
If $V_c\ne 0$ in the $N=Z$ system, $\delta R_M < 0$ is obtained,
owing to the second term in the parenthesis of Eq.(\ref{result}). In fact,
the negative $\delta R$ is observed in $^{40}$Ca in the analysis of Ref.\cite{kss}, as expected.
Fourth, the second term in the parenthesis expresses
the conventional understanding that the Coulomb
energy favors the increasing $R_p$, whereas $J$ prevents from the increasing $\delta R$\cite{bm}.
Fifth, the third term appears as order of $V_c^2$ which must be
smaller than the first two terms. It is proportional to $L$ which decreases the nucleon density
as in Eq.(\ref{dex}) through $\beta_\tau$ of Eq.(\ref{be}),
and inversely proportional to $K$ for playing a role against the change of
the nucleon density. Finally, it seems to be hard to find the relationship of Eq.({\ref{result}})
to the previous formula based on the droplet model\cite{roca1}.

\section{Finite nuclei in the mean field models}\label{finite}

\subsection{Fermi-type distribution}\label{fermi}

The Fermi-type function is  widely used for discussions of the nucleon distribution\cite{bm}.
It is written as
\begin{equation}
 \rho_\tau(r)
  = \rho_{0,\tau}\frac{1}{1+\exp\left((r-R_{{\rm d},\tau})/a_{{\rm d},\tau}\right)}.
\label{fermi}
\end{equation}
From the above equation, one has\cite{ks2,bm} 
\begin{equation}
R^2_\tau \approx \frac{3}{5}\left(\frac{3N_\tau}{4\pi\rho_{0,\tau}}\right)^{2/3}
 +\pi^2 a^2_{{\rm d},\tau},\label{fmsr}
 \end{equation}
which provides $\delta R$ as
\begin{equation}
\delta R\approx \delta R_0+\delta R_a   \label{dr}
\end{equation}
with
\begin{equation}
\delta R_0 = \left((1-\epsilon)^{-1/3}-1)\right)R_p\,,
 \qquad \epsilon
 =1-\frac{Z}{N}\frac{ \rho_{0,n}}{ \rho_{0,p}}\,,\label{dr0}
\end{equation}
and
\begin{equation}
\delta R_a=\frac{\pi^2}{2}\frac{a^2_{{\rm d},n}-a^2_{{\rm d},p}}{R_p}.\label{dra}
\end{equation}
In fact, the first term of Eq.(\ref{dr}) includes a small contribution from
the diffuseness parameter $a_{{\rm d},\tau}$
which is order of $(a_{{\rm d},\tau}/R_{{\rm d},\tau})^2$.
It is more important for the present purpose, however,
that if Eqs.(\ref{r}) and (\ref{dex}) with Eqs.(\ref{al}) and (\ref{be})
is used in Eq.(\ref{dr0}) whose $\rho_{0,\tau}$ is replaced by $\rho_\tau$
of SINM, the same expression as Eq.(\ref{result}) is obtained for $\delta R_0$. 
This fact is because in SINM, $\epsilon$ is expanded up to order of $V_c^2$ as
\begin{equation}
(1-\epsilon)^{-1/3}\approx (1+\frac{2I}{3})\left(1-\frac{2}{3}\alpha_nV_c
			    +\frac{2}{9}\alpha_n^2V_c^2\right),
\end{equation}
while $R_p$ is written as
\begin{equation}
R_p=\sqrt{\frac{3}{5}}\left(\frac{9\pi A}{8}\right)^{1/3}(1-\frac{I}{3})
 \left(1-\frac{1}{3}\alpha_pV_c+\frac{2}{9}\alpha_p^2V_c^2-\frac{1}{6}\beta_pV_c^2\right).
 \end{equation}
Thus,
$\delta R_0$ dominated by $R_{{\rm d},\tau}$ comes mainly
from the flat part of the Fermi-type distribution, corresponding to $\delta R_M$,
while $\delta R_a$ is responsible for its diffuseness.

Ref.\cite{ks2} has determined the values of $R_{{\rm d},\tau}$ and $a_{{\rm d},\tau}$
by the minimizing the volume integral of the square of the difference
between Eq.(\ref{fermi})
and $\rho_\tau(r)$ calculated in the MF models under the normalization,
$N_\tau=\int d\vct{r}\rho_\tau (r)$.
It has been shown that the neutron and proton distributions
in $^{208}$Pb calculated
in the RMF and SMF models are well reproduced
by the Fermi-type functions. Table \ref{table_msr} taken from Ref.\cite{ks2}
compares the average values of the root msr's obtained in the RMF and SHF models
to those of Eq.(\ref{fmsr}) from the approximated distributions with Fermi-type functions.
The former are listed as MF, and the latter as Eq.(\ref{fmsr}) in the table.
Since in most of the MF models, the value of $R_p$ is fixed by the experimental value
of the root msr of the charge density($R_c$) from electron scattering,
the framework-dependence of $\delta R$ is owing mainly to $R_n$.
Ref.\cite{ks2} has also shown that the values of $\rho_{0,\tau}$ obtained for $^{208}$Pb
is almost equal to those of the corresponding SINM as listed in Table \ref{table_dr}.
The nucleon density $\rho_{\tau}$ for SINM
in Table \ref{table_dr} has been obtained by the coupled equations from Eq.(\ref{me})
using $\eden$ in the RMF and SMF models\cite{ks2}.
This fact implies that the structure of $\delta R_0$ of $^{208}$Pb is able to be
discussed using $\delta R_M$ of SINM.

In the next subsection, neglecting $\delta R_a$ for a while,
the structure of $\delta R_0$ will be explored by using the NST formula in Eq.(\ref{result}).

\begin{table}
\begingroup
\renewcommand{\arraystretch}{1.2}
{\setlength{\tabcolsep}{4pt}
\hspace{4cm}
\begin{tabular}{|c|c|c|c|c|} \hline
\multicolumn{2}{|c|}{}       &
$R_n$  &
$R_p$  &
$\delta R$  \\ \hline
Rel   & MF                & 5.749 & 5.466 & 0.283 \\ \cline{2-5}
      &Eq.(\ref{fmsr}) & 5.728 & 5.451 & 0.277 \\ \hline
Non   &MF                 & 5.617 & 5.455 & 0.161 \\ \cline{2-5}
      &Eq.(\ref{fmsr}) &5.629&5.460&0.169   \\ \hline 
\end{tabular}
}
\endgroup
\caption{The mean values of the root msr's of the neutron($R_n$) and proton($R_p$) distributions 
and their difference $(\delta R=R_n-R_p)$ obtained for $^{208}$Pb\cite{ks2}. 
MF indicates the mean values in the relativistic(Rel) and non-relativistic(Non) mean-field models,
while
Eq.(\ref{fmsr}) stands for the equation number in the text used for the calculations of $R_\tau$
and $\delta R$ given in its row. All the values are listed in units of fm. For details,
see the text.}
\label{table_msr}
\end{table}

\begin{table}
\begingroup
\renewcommand{\arraystretch}{1.2}
{\setlength{\tabcolsep}{4pt}
\hspace{2.5cm}
\begin{tabular}{|c|c|c|c|c|c|c|c|} \hline
\multicolumn{2}{|c|}{}     &
$\rho_{0,\tau}$ &
$a_{{\rm den},\tau}$   &
$R_{{\rm den},\tau}$  &
$\epsilon$&
matter $\rho_{\tau}$&
matter $\epsilon$ \\ \hline 
Rel   & $n$  & $0.0860$ & $0.553$& $6.903$&$0.1044$ &$0.0832$ &$0.1138$\\ \cline{2-5}\cline{7-7}
      & $p$  & $0.0625$ & $0.454$& $6.692$&$$ &  $0.0611$&       \\ \hline
Non   & $n$  & $0.0911$ & $0.554$ &$6.766$ &$0.0556$ &$0.0903$&$0.0567$\\ \cline{2-5}\cline{7-7}
      & $p$  & $0.0628$ & $0.475$ &$6.672$ & $$ &$0.0623$&       \\ \hline 
\end{tabular}
}
\endgroup
\caption{The mean values of the parameters for the Fermi-type neutron(n) and proton(p) distributions, and those of corresponding nuclear matter\cite{ks2}.
The former is obtained by approximating the densities in the relativistic and non-relativistic
mean-field models for $^{208}$Pb, and the latter is determined by the HVH theorem
 for asymmetric  nuclear matter\cite{ks2}.
 The values of $\rho_{0,\tau}$ are given in units of fm$^{-3}$
and those of $a_{{\rm den},\tau}$ and $R_{{\rm den},\tau}$ are in fm.
 For the definition of $\epsilon$, see the text.}
\label{table_dr}
\end{table}

\subsection{The neutron-skin thickness $\delta R_0$}\label{I}

Eq.(\ref{result}) shows that $\delta R_M$ is independent of the detail of interaction
parameters in the MF models, regardless whether they are RMF or SMF.
Let us examine how Eq.(\ref{result}) is satisfied in the various MF models,
by using the same RMF and SMF models as in Ref.\cite{kss} and \cite{ks2},
since their conclusions are related to the present paper.

The MF models in Refs.\cite{kss} and \cite{ks2} are arbitrarily chosen among
many models
available at present\cite{stone}, as 
1 .L2\cite{sw1}, 2. NLB\cite{sw1}, 3. NLC\cite{sw1}, 4. NL1\cite{nl1}, 5. NL3\cite{nl3},
6. NL-SH\cite{nlsh}, 7. NL-Z\cite{nlz}, 8. NL-S\cite{nls}, 9. NL3II\cite{nl3}, 10. TM1\cite{tm1}
and 11. FSU\cite{fsu} for the RMF models,
and 1. SKI\cite{sk1}, 2. SKII\cite{sk1}, 3. SKIII\cite{sk3},
4. SKIV\cite{sk3}, 5. SkM$^\ast$\cite{skm}, 6. SLy4\cite{sly4}, 7. T6\cite{st6}, 8. SGII\cite{sg2}
and 9. Ska\cite{ska} for the SMF ones.
The above numbering of the models is according to Refs.\cite{kss,ks2}, and 
will be used throughout the present paper.  Note that in the present paper, the words of
the RMF and SMF models indicate the above ones.

Table \ref{table_rel} and \ref{table_non} show the values of
the quantities related to Eq.(\ref{result}) in the MF models.
Those of $k_{\rm F}$, $J$, $L$, and $K$ are taken from the references for the
models listed in the first column.
The values of
$\delta R$ are obtained by the RMF and SMF models, 
and $\delta R_0$ are calculated with the use of the Fermi-type
distribution corresponding to each model in Ref.\cite{ks2}
The values of $V_c$ 
are determined so that Eq.(\ref{result}) reproduces the value of $\delta R_0$.
It is seen that they are almost equal to the value of $22.144$ MeV used
in Ref.\cite{ks2}. This fact implies that Eq.(\ref{result}) reproduces well the values
of $\delta R_0$ in both the RMF and SMF models,
and that the following discussions are almost independent of the details
on the value of $V_c$. 
In the last column of the tables are listed
the values of the term related to $L$ in Eq.(\ref{result}),
\begin{equation}
\delta R_L
 =\sqrt{\frac{3}{5}}\frac{2}{3}\left(\frac{9\pi A}{8}\right)^{1/3}\frac{1}{k_{\rm F}}
\frac{I}{24}\left(\frac{3L}{2K}
	  \right)\left(\frac{V_c}{J}\right)^2,\label{rl}
\end{equation}
which depends not only on $L$, but also on $K$ and $J$.
As shown in the parentheses in the last column, $\delta R_L$ contributes to $\delta R_0$ by
less than $10\%$, but it is not negligible.
For example, the value of $J$ in  SKII(2) is larger than that in Ska(9), while the value
of $\delta R_0$ in the former is smaller than that in the latter, owing to the contributions
from $\delta R_L$, as in Table \ref{table_non}.

Tables \ref{table_rel} and \ref{table_non} show that the $\delta R_0$ explains more than
a half of $\delta R$ in most of the models.
Since the values of the $O(V_c^2)$-term in Eq.(\ref{result}) are small,
$\delta R_0$ is dominated by the first two terms depending on $k_{\rm F}$, $I$, $V_c$ and $J$.
Among them, the values of the first three quantities are almost the same in all the models,
the model-dependence of $\delta R_0$ is mainly due to $J$ in both RMF and SMF models.  
In Tables \ref{table_rel} and \ref{table_non}, the values of $J$ in the RMF models are larger
than those in the SMF models. This fact results in the larger $\delta R_0$
in the RMF models, compared with those in the SMF ones, as shown in Ref.\cite{kss}.  

It is concluded that $\delta R_0$ in the MF models is well explained by Eq.(\ref{result}).
Provided that the value of $k_{\rm F}$ and $V_c$ are almost the same in the MF models,
$\delta R_0$ depends mainly on $J$, and on a small contribution from $L$ and $K$. 
About a half of $\delta R$ comes from $\delta R_a$, which are about $0.090\pm 0.015$ fm
in the RMF models, while $0.070\pm 0.010$ fm in the SMF models, as discussed later.

\begin{table}
\begingroup
\renewcommand{\arraystretch}{1.2}
{\setlength{\tabcolsep}{4pt}
\hspace{0cm}
\begin{tabular}{|l|c|c|c|c|c|c|c|c|} \hline
Model&
$k_{\rm F}$ &
$J$&
$L$&
$K$&
$\delta R$&
$\delta R_0$&
$V_c$&
$\delta R_L\times 10^2$\\ \hline 
1 L2    & $1.3003$  & $34.981$ & $115.530$& $546.826$ &$0.27568$ &$0.19183$ &$22.435$&$0.4118(2.15\%)$\\ \hline
2 NLB   & $1.3001$  & $35.014$ & $108.259$& $421.024$ &$0.26423$ &$0.18137$ &$22.916$&$0.5220(2.88\%)$\\ \hline
3 NLC   & $1.3002$  & $35.021$ & $107.969$& $224.461$ &$0.26312$ &$0.17692$ &$23.290$&$1.0082(5.70\%)$\\ \hline
4 NL1   & $1.3098$  & $43.470$ & $140.099$& $211.122$ &$0.32119$ &$0.21358$ &$27.132$&$1.2162(5.69\%)$\\ \hline
5 NL3   & $1.2994$  & $37.398$ & $118.527$& $271.532$ &$0.28056$ &$0.19065$ &$24.234$&$0.8692(4.56\%)$\\ \hline
6 NL-SH & $1.2930$  & $36.127$ & $113.685$& $355.647$ &$0.26592$ &$0.18415$ &$23.623$&$0.6513(3.54\%)$\\ \hline
7 NL-Z  & $1.3066$  & $41.677$ & $133.731$& $172.713$ &$0.30977$ &$0.20609$ &$26.516$&$1.4783(7.17\%)$\\ \hline
8 NL-S  & $1.3046$  & $42.070$ & $131.588$& $262.944$ &$0.30637$ &$0.20640$ &$26.507$&$0.9384(4.55\%)$\\ \hline
9 NL3II & $1.3021$  & $37.701$ & $119.706$& $271.724$ &$0.28224$ &$0.19167$ &$24.372$&$0.8712(4.55\%)$\\ \hline
10 TM1  & $1.2906$  & $36.892$ & $110.794$& $281.159$ &$0.27072$ &$0.18417$ &$24.202$&$0.8096(4.40\%)$\\ \hline
11 FSU  & $1.2994$  & $32.561$ & $60.438$ & $229.538$ &$0.20691$ &$0.11804$ &$23.703$&$0.6617(5.61\%)$\\ \hline 
\end{tabular}
}
\endgroup
\caption{The values of the Fermi momentum($k_{\rm F}$), the asymmetry energy coefficient($J$),
 the slope($L$) of the asymmetry energy, the incompressibility coefficient($K$),
 the difference between the root mean square radii of the neutron and proton
 distributions($\delta R$), $\delta R_0$ in Eq.(\ref{dr}), the Coulomb energy($V_c$),
 and $\delta R_L$ in Eq.(\ref{rl}) 
 given by the relativistic MF models in the  first column.
The units are fm$^{-1}$ for $k_{\rm F}$, MeV for $J$, $L$, $K$, and $V_c$, and fm for $\delta R$,
$\delta R_0$ and $\delta R_L$
The ratio of $\delta R_L$ to $\delta R_0$ is listed in the parenthesis of the last column.
 For details, see the text.}
\label{table_rel}
\end{table}

\begin{table}
\begingroup
\renewcommand{\arraystretch}{1.2}
{\setlength{\tabcolsep}{4pt}
\hspace{2mm}
\begin{tabular}{|l|c|c|c|c|c|c|c|c|} \hline
Model&
$k_{\rm F}$&
$J$&
$L$&
$K$&
$\delta R$&
$\delta R_0$ &
$V_c$&
$\delta R_{\rm L}\times 10^2$ \\ \hline 
1 SKI        & $1.3200$  & $29.242$ & $1.219$&  $370.380$ &$0.11410$ &$0.04955$ &$23.319$&$0.09770(1.97\%)$\\ \hline
2 SKII       & $1.2999$  & $34.158$ & $50.024$& $341.404$ &$0.19435$ &$0.12048$ &$24.653$&$0.3618(3.00\%)$\\ \hline
3 SKIII      & $1.2908$  & $28.161$ & $9.907$&  $355.368$ &$0.12483$ &$0.05592$ &$22.313$&$0.08355(1.49\%)$\\ \hline
4 SKIV       & $1.3073$  & $31.218$ & $63.496$& $324.551$ &$0.19185$ &$0.12914$ &$22.237$&$0.4679(3.62\%)$\\ \hline
5 SkM$^\ast$ & $1.3338$  & $30.033$ & $45.776$& $216.609$ &$0.16893$ &$0.09181$ &$22.643$&$0.5549(6.04\%)$\\ \hline
6 SLy4       & $1.3317$  & $32.004$ & $45.960$& $229.911$ &$0.15979$ &$0.08730$ &$24.294$&$0.5330(6.11\%)$\\ \hline
7 ST6        & $1.3354$  & $29.966$ & $30.853$& $235.947$ &$0.15062$ &$0.07008$ &$23.382$&$0.3642(5.20\%)$\\ \hline
8 SGII       & $1.3284$  & $26.830$ & $37.629$& $214.649$ &$0.13497$ &$0.06694$ &$20.996$&$0.4980(7.44\%)$\\ \hline
9 Ska        & $1.3200$  & $32.910$ & $74.623$& $263.155$ &$0.21137$ &$0.13833$ &$23.119$&$0.6532(4.72\%)$\\ \hline
\end{tabular}
}
\endgroup
\caption{The values of the Fermi momentum($k_{\rm F}$), the asymmetry energy coefficient($J$),
 the slope($L$) of the asymmetry energy, the incompressibility coefficient($K$),
 the difference between the root mean square radii of the neutron and proton
 distributions($\delta R$), $\delta R_0$ in Eq.(\ref{dr}), the Coulomb energy($V_c$),
 and $\delta R_L$ in Eq.(\ref{rl}) 
 given by the non-relativistic MF models in the first column.
The units are fm$^{-1}$ for $k_{\rm F}$, MeV for $J$, $L$, $K$, and $V_c$, and fm for $\delta R$,
$\delta R_0$ and $\delta R_L$.
The ratio of $\delta R_L$ to $\delta R_0$ is listed in the parenthesis of the last column.
 For details, see the text.}
\label{table_non}
\end{table}

\subsection{Least squares analysis}\label{lsa}

The NST formula in Eq.(\ref{result}) shows that $\delta R_M$ depends
on several quantities like $J$, $L$, and $K$. It has been shown that the values
of $\delta R_0$ of the RMF and SMF models also approximately satisfy the NST formula.
In this subsection, it is investigated on the basis of Eq.(\ref{result})
how the relationship appears in the least square lines(LSL)
between $\delta R$ and these macroscopic quantities
calculated in the various MF models.
In particular, the linear correlation between $\delta R$ and $L$
observed in the previous papers\cite{thi,jlab1,jlab2,rein0,roca1,roca2,roca3} will be
investigated in detail,
since Eq.(\ref{result}) shows that the coefficient of $L$ is not a constant,
but is given model-dependently on $J$ and $K$, in addition to the fact that
the $L$-term contributes to $\delta R_0$ by less than $10\%$ only, 
as shown in Tables \ref{table_rel} and \ref{table_non}.

For understanding what one learns from LSL by LSA, it is instructive to refer to the results
of Ref.\cite{kss}. It has recently estimated the values
of $R_n^2$ from the small component of the fourth moment of the nuclear
charge distribution($Q^4_c$) 
in $^{40}$Ca, $^{48}$Ca and $^{208}$Pb, by LSA using electron-scattering data.
At the beginning of this subsection, their LSL's are briefly reviewed,
in a different point of view from that in Ref.\cite{kss}.

\subsubsection{The moments of the neutron and proton distributions}\label{mom}

The msr of the nuclear charge density($R^2_c$) is derived almost model-independently
from electron scattering data, since the relationship between the charge density
and the scattering cross section is well understood theoretically.
The value of $R^2_p$ is also well determined by using the established 
relational formula between $R^2_p$ and $R^2_c$
as $R^2_c=R^2_p +\Delta_p$ \cite{ks1}, where $\Delta_p$ is given by the terms of
the nucleon size and those depending weakly on the nuclear structure.
The contribution of $\Delta_p$ to $R^2_c$
is less than $1\%$\cite{ks1,kss}. 
In fact, however, nuclear phenomenological models are not constructed so as to
satisfy exactly the relationship in fixing the values of the free interaction
parameters\cite{nl3,sly4}. As a result, for example, the MF models
provide different values of $R^2_p$.
Nevertheless, in plotting those values in the ($R^2_p-R^2_c$)-plane,
the equation of LSL as $R^2_c=a_pR^2_p+b_p$ has
a small standard deviation, reflecting the fact that
each element $(R^2_p, R^2_c)$ of the set satisfies almost
the relationship of $R^2_c=R^2_p +\Delta_p$ \cite{kss}.
Hence, when the experimental value of $R^2_c$
is given in the LSL-equation, the value of $R^2_p$ is determined
for the used model-framework.
Note that it is not necessary to have the same LSL-equation for the RMF
and SMF models, mainly because the values of $\Delta_p$ are different
from each other\cite{kss}.
Moreover, the values of the constant $a_p$ and $b_p$ depend on the model-dependent
distribution of the points ($R^2_p, R^2_c$) in the ($R^2_p-R^2_c$)-plane.
The reference formula as $R^2_c=R^2_p +\Delta_p$, however, 
is necessary for distinguishing the correlation
between the two variables from a kind of spurious ones, as explained later.

Main purpose of Ref.\cite{kss} is to determine the value of the msr 
of the point neutron distributions in the MF models,
using experimental data
from conventional electron-scattering\cite{vries}.
They have analyzed the LSL-equations for the fourth moment($Q^4_c$) of the charge
distribution and its small component with $R^2_n$.
Hence, it is useful to understand their results for exploring the relationship
between $\delta R$ and the small component with $L$. 
The equations of Ref.\cite{kss} are simpler in the relativistic framework
than in the non-relativistic one, so that the former will be used for discussions below.
The latter has several relativistic corrections\cite{kss}
which are not essential for the present discussions.

The fourth moment in the relativistic framework
is described  using the same notations as in Ref.\cite{kss} as,
\begin{equation}
Q^4_c= Q^4_{cp}-Q^4_{cn}\,,\label{4thm}
\end{equation} 
where the fourth moments of the proton($ Q^4_{cp}$) and neutron($Q^4_{cn}$)
charge distributions are given,
respectively, by
\begin{align}
Q^4_{cp}&= Q_{p}^4+Q_{2p}+Q_{2W_p}+Q_{4W_p}+(Q_4)_p\,, \label{qcp}\\
Q^4_{cn}&= Q_{2n}+Q_{2W_n}+Q_{4W_n}+(Q_4)_n\,.\label{qcn}
\end{align}
The main component of $Q^4_c$ is $Q^4_{cp}$, whose value calculated in the MF models
exceeds the experimental one of $Q^4_c$.
The small component $Q^4_{cn}$ yields a negative contribution by
about $3\%$ of $Q^4_c$ to reproduce the experimental value\cite{ks1,kss}.
The main component of $Q^4_{cp}$ is the fourth moment of the point proton distribution($Q^4_p$).
The contribution of the term $Q_{2p}$ with $R^2_p$ to $Q^4_c$
is less than $10\%$ of $Q^4_c$, while the one from $Q_{2n}$ with $R^2_n$
about $2\%$\cite{kss}. They depend on $R^2_p$ and $R^2_n$ as
\begin{equation}
Q_{2p}=\frac{10}{3}r_p^2R^2_p,\qquad
 Q_{2n}=-\frac{10}{3}(r_+^2-r_-^2) R^2_n\frac{N}{Z}. \label{q2n}
\end{equation}
Note that the coefficients of $R^2_p$ and $R^2_n$ are given by the proton($r_p^2$) and
neutron($r^2_n=r_+^2-r_-^2$) msr's, respectively, and are independent
of nuclear structure parameters.
The other terms in Eqs.(\ref{qcp}) and (\ref{qcn}) yield small contributions,
which are from the spin-orbit density\cite{ks1,ks0}
and from the fourth moment of proton($(Q_4)_p =5r_p^4/2,$)
and neutron($(Q_4)_n =-5(r_+^4-r_-^4) N/2Z$) charge density\cite{ks1}.

The experimental value of $Q^4_c$ in $^{208}$Pb has been determined precisely
through electron scattering as $1171.981(17.627)$ fm$^4$ with the experimental
error in the parenthesis\cite{vries}.
Using the same RMF models as in \S\ref{I}, Ref.\cite{kss} has obtained the
the LSL-equation for $Q^4_{cp}$ and $Q^4_c$ as
\begin{equation}
Q_c^4=0.9972Q^4_{cp}-29.5610\,, (\sigma=0.3790) \label{42cpa},
\end{equation}
where $\sigma$ indicates the standard deviation of the LSL, and
the numbers are written so that each moment is given with its own unit in fm,
as $Q^4_c$ in fm$^4$.
In all the following equations, the units will be given in the same way.
Inserting the experimental value of $Q^4_c$ into the above equation, one has\cite{kss}
\begin{equation}
Q^4_{cp}=1204.875(18.056).\label{qcpr}
\end{equation}
The LSL-equation for $Q^4_{cn}$ and $R^2_n$ is given by Ref.\cite{kss} as
\begin{equation}
Q^4_{cn}=0.2689R^2_n+24.0364(\sigma=0.3599).\label{qcna}
\end{equation}
In using Eq.(\ref{4thm}) and the value of $Q^4_{cp}$ in Eq.(\ref{qcpr}),
the value of $Q^4_{cn}$ is obtained as $Q^4_{cn}=32.895(0.429)$ fm$^4$.
This value and Eq.(\ref{qcna}) determine the value of $R^2_n$ as
\begin{equation}
R^2_n=32.943(2.934).\label{nv}
\end{equation}
It is remarkable that the small contribution from $Q^4_{cn}$ to $Q^4_c$
makes it possible to determine the value of $R^2_n$.

On the one hand, with respect to $R^2_p$ and $Q^4_{cp}$, Ref.\cite{kss} has obtained 
\begin{equation}
Q^4_{cp}=81.6203R^2_p-1230.9324\,,(\sigma=2.5499)\label{42ppa2}, 
\end{equation}
This and Eq.(\ref{qcpr}) yields
\begin{equation}
R^2_p=29.843(0.252).\label{pv2}
\end{equation}
On the other hand, the correlation of $R^2_p $ with $R^2_c$ has also been examined.
As mentioned before, the relationship between $R^2_c$ and $R^2_P$
in the relativistic framework is written as\cite{ks1}
\begin{equation}
R^2_c =R^2_p + r_p^2 +r^2_n\frac{N}{Z}+R^2_{W_p}+R^2_{W_n} \frac{N}{Z},
\label{msr}
\end{equation}
where
the last two terms stand for small contributions from the
spin-orbit density\cite{ks1,ks0}.
Thus, $R^2_c$ is dominated by $R^2_p$ and is independent of $R^2_n$.
Ref.\cite{kss} has obtained the LSL-equation as
\begin{equation}
R^2_c=1.0024R^2_p+0.4791\,, (\sigma=0.0015). \label{rcp}
\end{equation}
Giving the experimental value of $R^2_c$
to be $R^2_c=30.283(0.154)$ fm$^2$ in the left-hand side\cite{vries},
one can determine the value of $R^2_p$ as
\begin{equation}
R^2_p=29.733(0.155).\label{pv}
\end{equation}
The above value is almost the same as that in Eq.(\ref{pv2}) obtained
from the experimental value of $Q^4_c$.
In principle, $Q^4_c$ and $R^2_c$ are independent of each other as in Eqs.(\ref{4thm})
and (\ref{msr}).
This agreement may be related to the fact that the proton distributions
in the MF models are well approximated by the Fermi-type function\cite{ks2}
where all the moments correlate with each other\cite{bm}.

Similar LSL's have been obtained by LSA for the SMF models in Ref.\cite{kss},
and the experimental values of $R_c^2$ and $Q^4_c$ are reproduced consistently,
as in the RMF models.
The value of $R^2_n$ is, however, smaller by about $0.1$ fm than that in the RMF
models\cite{kss}. This result is because LSA is not a method to deduce the experimental value,
but provides the value of $R^2_n$ allowed in each model-framework.
The reason of the $0.1$ fm difference has been shown to be understood
according to the HVH theorem in Ref.\cite{ks2}.

For discussing the correlation between $\delta R_0$
and its small component with $L$ below,
one more comment from Ref.\cite{kss} is required.
Eq.(\ref{result}) and the analysis in Ref.\cite{ks2} have shown
that $R^2_n$ and $R^2_p$
are not independent variables owing to HVH theorem in the MF framework.
In such a case, one obtains LSL for a kind of the
``spurious'' correlation which is not expected from the reference formula
like Eq.(\ref{msr}).
In fact, Ref.\cite{kss} has obtained the well-defined LSL
for the correlation between $R^2_n$ and $R^2_c$ as 
\begin{equation}
R^2_c=0.5154R^2_n+13.4036\,,(\sigma=0.2170), \label{rcn}
\end{equation}
even though there is no contribution from $R^2_n$ to $R^2_c$ as shown in Eq.(\ref{msr}).
The above equation together with Eq.(\ref{rcp}) provides the relationship as
\begin{equation}
 R^2_n=1.9449R^2_p -25.0766.\label{pnc2}
\end{equation}
Ref.\cite{kss} has also estimated the value of $R^2_n$ directly by using $Q^4_c$.
They have obtained LSL's in the ($R^2_p-Q^4_c$)- and ($R^2_n-Q^4_c$)-planes
for the values calculated in the RMF models as
\begin{align}
Q^4_c&=81.4556R^2_p-1258.9107\,,(\sigma =2.3239)\,,\label{qcpp}\\ 
Q^4_c&=43.10936R^2_n-249.0948\,,(\sigma=16.9393).\label{qcnn}
\end{align}
On the one hand, when using Eqs.(\ref{42cpa}) and (\ref{42ppa2}), one obtains
\begin{equation}
Q^4_c=81.3918R^2_p-1257.0468,\label{2step}
 \end{equation}
which is almost the same as Eq.(\ref{qcpp}).
The experimental value of $Q^4_c$ provides the
value of $R^2_p$ to be $29.843(0.245)$ fm$^{2}$ from Eq.(\ref{qcpp})
which is almost the same 
as $R^2_p=29.733(0.155)$ fm$^{2}$ in Eq.(\ref{pv}) through the experimental value
of $R^2_c$, in spite of the fact that the term $Q_{2p}$ with $R^2_p$
contributes to $Q^4_{c}$ by less than $10\%$ in Eq.(\ref{qcp}).
On the other hand, Eq.(\ref{qcnn}) seems to be inconsistent with Eq.(\ref{4thm}),
since the contribution from $Q^4_{cn}$ with $R^2_n$ to $Q^4_c$ is negative,
while the coefficient of $R^2_n$ in Eq.(\ref{qcnn}) is positive.
This result, however, owing to the spurious correlation as in Eq.(\ref{rcn}).
Since the contribution of the term with $R^2_n$ to $Q^4_c$ is small, the correlation between
$R^2_n$ and the main term $Q^4_{cp}$ appears in Eq.(\ref{qcnn}).
Indeed, equating Eq.(\ref{qcpp}) to Eq.(\ref{qcnn}) yields
\begin{equation}
R^2_n=1.8895R^2_p-23.4245.\label{pnc}
\end{equation}
which is almost equal to Eq.(\ref{pnc2}).
Thus, Eqs.(\ref{rcn}) and (\ref{qcnn}) are the LSL-equations for
a kind of spurious correlations,
owing to Eq.(\ref{pnc2}) or (\ref{pnc}).

Eqs.(\ref{pnc2}) and (\ref{pnc}) are understood as a relationship constrained
through Eq.(\ref{result}) due to the HVH theorem in the MF models.
Hence, Eqs.(\ref{rcn}) and (\ref{qcnn}) do not
contradict other equations. For example, the value of $R^2_n$ 
in Eq.(\ref{nv}) is 32.943(2.934), while inserting the experimental values
of $R^2_c$ in Eq.(\ref{rcn}) and of $Q^4_c$ in Eq.(\ref{qcnn}), Ref.\cite{kss} has
obtained $R^2_n=32.752(0.720)$ and $32.964(0.802)$ in units of fm$^2$, respectively.

To summarize \S 3.3.1,
in spite of the fact that the contribution from $R^2_n$ to $Q^4_c$ is
less than a few $\%$, the value of $R^2_n$ is determined for each model-framework
from the experimental value of $Q^4_c$ .
It is necessary for the determination
to have a well-defined LSL as in Eq.(\ref{qcna}) between $R^2_n$ and
the small component $Q^4_{cn}$ whose value is estimated from experiment.
Since $R^2_n$ is not independent of $R^2_p$ owing to the HVH theorem,
there is a kind of spurious correlations between $R^2_n$ and $R^2_c$, and
between $R^2_n$ and $Q^4_c$.
 
Now bearing in mind the above discussions, correlations of $\delta R_0$
with macroscopic parameters
are examined by LSA.

\subsubsection{Correlation of $\delta R_0$ with nuclear matter parameters}\label{lsll}

For the LSA of the moments, Eq.(\ref{4thm}) has been necessary as a reference formula.
In the same way, it is reasonable to use  Eq.(\ref{result}) as the reference formula 
for LSA of the correlation between $\delta R_0$ with macroscopic parameters.
Tables \ref{table_rel} and \ref{table_non} show
that $\delta R_0$ is dominated
by $I$ and $1/J$ in the parenthesis in Eq.(\ref{result}).
Hence, the correlation of $1/J$ should be examined first.

\begin{figure}[ht]
\centering{%
\includegraphics[scale=1]{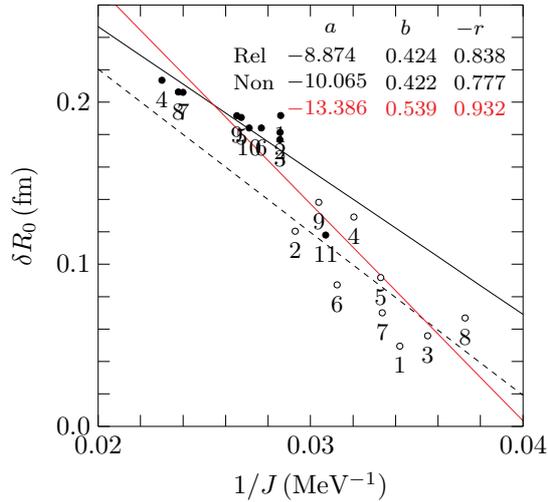}
}
\caption{
The correlation between $1/J$ and $\delta R_0$
in $^{208}$Pb.
Here, $J$ denotes the asymmetry energy coefficient, and $\delta R_0$ the neutron-skin thickness
without the diffuseness parts of the neutron and proton densities.
The closed circles show the values of $\delta R_0$ calculated in the RMF models
using the point neutron and proton distributions approximated by the Fermi-type functions,
while the open circles in the SMF models. Each circle is accompanied
by the number which indicates the used model specified in the text.
The black solid and dashed lines show the least square lines(LSL) for the closed and open circles,
respectively, while the red one is obtained by taking account of all the circles.
The values of the slope($a$), the intercept($b$) and the correlation coefficient($r$) of the LSL's
are written inside of the figure. For details, see the text.
}
\label{figure_RO-J}
\end{figure}

In Fig.\ref{figure_RO-J} is
shown the correlation between $1/J$ and $\delta R_0$ in the RMF and SMF
models cited in Ref.\cite{kss}.
The points of the RMF models are indicated by the closed circles, while those of
the SMF ones by the open circles. Each point is attached with the number showing
the used model which is listed in Tables \ref{table_rel} and \ref{table_non}.
The LSL, $\delta R_0=a/J+b$, is shown for the RMF by the solid line and for
SMF models by the dashed line, together with the values of the correlation coefficient $r$.
They are drown for the RMF and SMF models, separately, since
there is no specific reason why both frameworks should be analyzed together, as in the
case of ($R^2_n-Q^4_c$)-correlation in Ref.\cite{kss}.
The functions of $J(\rho)$ which provides $J=J(\rho_0)$
are given by different parameter sets in the two frameworks,
respectively, as
\begin{align}
J_{\rm rel}(\rho)&=\frac{k_{\rm F}^2}{6\sqrt{k^2_{\rm F}+M^ 2_\sigma}}+\frac{\rho}{2}\frac{g^2_\rho}
{m^2_\rho+2\lambda g^2_\rho V^2_\omega},\,\qquad  M_\sigma=M+V_\sigma, \label{rs2}
\\[4pt]
J_{\rm non}(\rho)&= c_k\left(\frac{\rho^{2/3}}{6M}
 +\frac{-3t_1x_1+t_2(5x_2+4)}{24}\rho^{5/3}\right) \nonumber \\[2pt]
&
\hphantom{=}
 -\frac{2x_0+1}{8}t_0\rho-\frac{2x_3+1}{48}t_3\rho^{\alpha+1},\,
 \quad (c_k= (3\pi^2/2)^{2/3}).\label{ns2} 
\end{align}
The notations in the above equations are the same as in Ref.\cite{ks2}.
Note that Eq.(\ref{rs2}) represents $J(\rho)$ of the RMF models
including FSU(11) in which two parameters are added to those,
for example, in NL3(5),
aiming to reduce the difference between $R^2_n$'s in the RMF
and SMF models\cite{fsu}.One of them is $\lambda$ which appears in Eq.(\ref{rs2}).
\color{black}
Indeed, the LSL's for the RMF and SMF models are well defined
and clearly separated from each other as seen in Fig.\ref{figure_RO-J}.
FSU(11) is seen to be the exception, whose point is 
almost on the SMF line,
because of $\lambda$ which makes the value of $J$ small in Eq.(\ref{rs2}). 

The value of $r$ in the SMF models is smaller than that in the RMF models.
This fact reflects partially that contributions from the component $\delta R_L$
in Eq.(\ref{result}) are small, but are not negligible,
and they are different in the two frameworks.
Table \ref{table_rel} shows $2.15\%$ to $7.17\%$ contribution from $\delta R_L$,
while Table \ref{table_non}  does $1.49\%$ to $7.44\%$.

If the LSL is calculated for the two frameworks together,
as in the previous papers\cite{roca1,roca2,roca3}, the red line is obtained
with $r=0.932$, in contrast to $r=0.838$ in the RMF models and $0.777$ in the SMF ones.
In analyzing results in the two frameworks together,
such an improvement is sometimes recognized later also, as far as the value of $r$
is concerned. One of the reasons of the improvement
is due to the definition of $r$.  It is given by $r^2=1-\sigma^2/(\Delta y)^2 $ with
$\Delta y=(\langle y^2\rangle-\langle y\rangle^2)^{1/2}$
and $n\sigma^2=\sum_{i=1}^n(y_i-ax_i-b)^2$ for the LSL-equation $y=ax+b$\cite{kss}.
When the value of $\sigma$ does not change so much in the LSL's of each framework
and in the LSL of both ones together,
but that of $\Delta y$ in the latter LSL becomes larger,
the value of $r^2$ in the latter LSA approaches to $1$ by the definition.
Since the relativistic and non-relativistic models
yield the calculated values located at a different region from each other
in the ($1/J-\delta R_0$) plane, the value of $\Delta y$ becomes larger
in the LSA for all the calculated values together.
In this sense, the improvement of the value of $r$
does not seem to have a physical meaning. 

As mentioned in \S\ref{intro}, the previous papers have pointed out the strong
correlation of $L$ with $\delta R$\cite{jlab1,jlab2,roca1,roca2,roca3}.
It may be explored next the correlation of $L$ with $\delta R_0$. 
The NST formula in Eq.(\ref{result}), however, has the $L$-dependence
in the small component of $\delta R_0$, together with $K$- and $J$-dependence.
It should be investigated how $\delta R_L$ correlates with $L$,
before discussing the ($L-\delta R_0$)-one, as in the case of the ($R^2_n-Q^4_{cn}$)-correlation.

\begin{figure}[ht]
\centering{%
\includegraphics[scale=1]{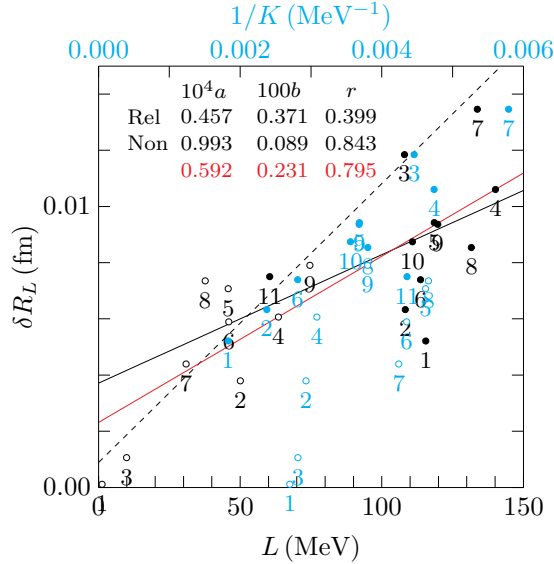}
}
\caption{The correlations between $L$ and $\delta R_L$(black),
and between $1/K$ and  $\delta R_L$(blue) in $^{208}$Pb. Here, $L$
and $K$ denote the slope of the asymmetry
energy and incompressibility coefficient, respectively, and $\delta R_L$
is defined by Eq.(\ref{rl}).
The closed black and blue circles show the values of ($L,\delta R_L$) and ($1/K,\delta R_L$)
in the RMF models,
while the open ones indicate those in the SMF models.  
Each circle is accompanied
by the number which represents the used model specified in the text.
The bottom scale-line is for $L$, and the top one for $1/K$.
The black solid and dashed lines show the least square lines(LSL) for the black closed and open circles,
respectively, while the red one is obtained by taking account of all the black circles.
The values of the slope($a$), the intercept($b$) and the correlation coefficient($r$) of the LSL's
are written inside of the figure.
For details, see the text.
}
\label{figure_RL-LK}
\end{figure}

\begin{figure}[ht]
\centering{%
\includegraphics[scale=1]{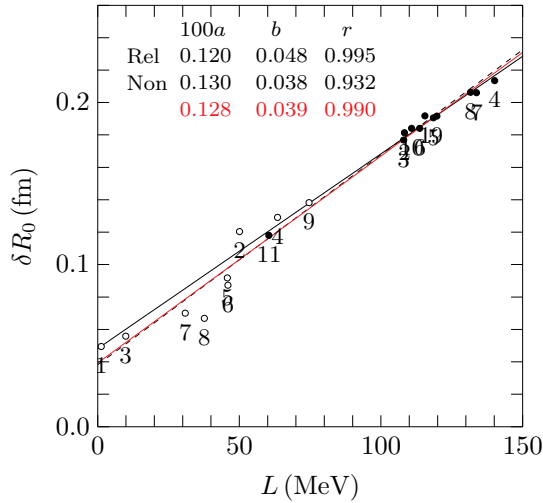}
}
\caption{
The correlation between $L$ and $\delta R_0$
in $^{208}$Pb.
Here, $L$ denotes the slope of the asymmetry energy, and $\delta R_0$ the neutron-skin thickness
without the diffuseness parts of the neutron and proton densities.
The closed circles show the values of $\delta R_0$ calculated in the RMF models
using the point neutron and proton distributions approximated by the Fermi-type functions,
while the open circles in the SMF models. Each circle is accompanied
by the number which indicates the used model specified in the text.
The black solid and dashed lines show the least square lines(LSL) for the closed and open circles,
respectively, while the red one is obtained by taking account of all the circles.
The values of the slope($a$), the intercept($b$) and the correlation coefficient($r$) of the LSL's
are written inside of the figure. For details, see the text.
}
\label{figure_RO-L}
\end{figure}

In Fig.\ref{figure_RL-LK} is shown the correlation between $L$ and $\delta R_L$
given in Eq.(\ref{rl}) whose values are calculated with $V_c=22.144$ MeV as in Ref.\cite{ks2}.
The filled and open black circles show the correlations of ($L-\delta R_L$) in the relativistic
and non-relativistic frameworks, respectively. The bottom scale-line is written for $L$
in units of MeV, and the vertical line for $\delta R_L$ in fm. 
It is seen that there is not a clear linear correlation in the RMF models.
The open circles for the SMF models increase gradually, with increasing $L$,
but as seen later in Fig.\ref{figure_RO-L}, the value of the slope of the LSL
is smaller than that in the ($L-\delta R_0$)-correlation by one order of magnitude,
and the order of the models according to the magnitude of $\delta R_L$ does not coincide
with that according to the magnitude of $\delta R_0$.
Thus, the small component $\delta R_L$ 
does not seem to play an essential role in the ($L-\delta R_0$) correlation. 

\begin{figure}[ht]
\centering{%
\includegraphics[scale=1]{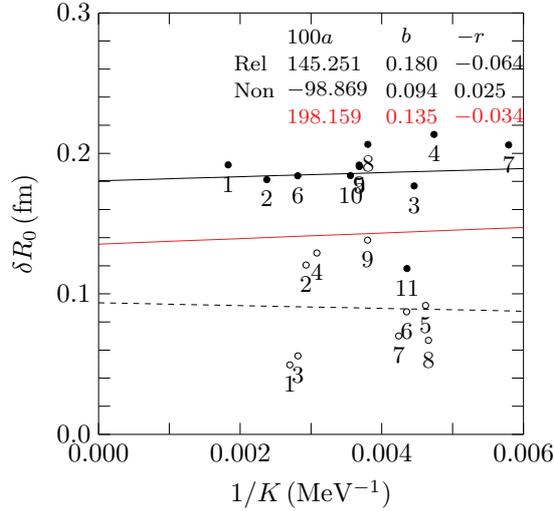}
}
\caption{
The correlation between $1/K$ and $\delta R_0$
in $^{208}$Pb.
Here, $K$ denotes the incompressibility coefficient, and $\delta R_0$ the neutron-skin thickness
without the diffuseness parts of the neutron and proton densities.
The closed circles show the values of $\delta R_0$ calculated in the RMF models
using the point neutron and proton distributions approximated by the Fermi-type functions,
while the open circles in the SMF models. Each circle is accompanied
by the number which indicates the used model specified in the text.
The black solid and dashed lines show the least square lines(LSL) for the closed and open circles,
respectively, while the red one is obtained by taking account of all the circles.
The values of the slope($a$), the intercept($b$) and the correlation coefficient($r$) of the LSL's
are written inside of the figure. 
For details, see the text.
}
\label{figure_RO-K}
\end{figure}

For reference, 
the correlation between $1/K$ and $\delta R_L$
is shown in Fig.\ref{figure_RL-LK}, where
the filled and open blue circles show the values of $(1/K, \delta R_L)$ in the RMF
and SMF models, respectively. The top line indicates the scale for $1/K$ in units
of MeV$^{-1}$. In contrary to the $(L-\delta R_L)$-correlations,
the closed blue points in the RMF models
show the increasing $\delta R_L$ with $1/K$,
whereas the open ones in the SMF models do not. Such a correlation in the RMF models,
however, will be seen later
to disappear in the ($1/K-\delta R_0$) correlation in Fig.\ref{figure_RO-K}.

In the analysis of the moments,
the small component $Q^4_{cn}$ of $Q^4_c$ plays an important role
for estimating the value of $R^2_n$.
In the case of the NST,
even if $\delta R_L$ is separated from the main part of $\delta R_0$
by utilizing the experimental value as $Q^4_{cn}$,
it may not be useful for estimating the value of $L$ or $K$.

Now, Fig.\ref{figure_RO-L} shows the correlation between $L$ and $\delta R_0$
in the same designation as in Fig.\ref{figure_RO-J}.
In spite of the fact that the contribution from $L$ to $\delta R_0$ is much smaller than
from $1/J$,
the linear correlation between them is more clearly seen
than that in Fig.\ref{figure_RO-J}.
If this LSL is obtained through $\delta R_L$, there would be the similar correlation
of $\delta R_0$ with $1/K$.

Fig.\ref{figure_RO-K} shows the correlations between $1/K$ and $\delta R_0$,
where the filled and open circles indicate the values of $(1/K, \delta R_0)$ in the RMF
and SMF models, respectively.
The bottom line indicates the scale of $1/K$ in units of MeV$^{-1}$.
It is seen that Fig.\ref{figure_RO-K} has no similarity to Fig.\ref{figure_RO-L},
in spite of the fact that $L$ and $1/K$ are in the same component of Eq.(\ref{result}).
Thus, the linear correlation of $L$ with $\delta R_0$
does not seem to reflect directly the structure of the NST formula
which is the reference formula.

The reason of the linear correlation of $L$ with $\delta R_0$ may be found 
in the analysis of the moments in \S \ref{mom}.
It has been pointed out that
there are spurious correlations in LSA which are not explained by the reference formulae.
The linear relationships of $R^2_n$ in the small component $Q^4_{cn}$
with $Q^4_c$, and with $R^2_c$ are spurious ones.
The spurious correlation between $R^2_n$ and
$Q^4_c$ appears through the main component $Q^4_{cp}$ which is related to $R^2_p$.
If the ($L-\delta R_0$)-correlation is spurious,
it may be due to the
the correlation between $L$ and the main component $1/J$
in the NST formula.
Such a correlation is expected, since $K$ is independent of $J$,
whereas $L$ is defined through $J$ as\cite{br,roca2}
\begin{equation}
L=3\rho\frac{\partial J(\rho)}{\partial \rho}|_{\rho=\rho_0}\label{L2},
\end{equation}
although there is no constraint on the $\rho$-dependence of $J(\rho)$
in the MF framework, and it is not necessary to have the dependence as in
Eqs.(\ref{rs2}) and (\ref{ns2}).

So far,
the proton and neutron distributions in the MF models are approximated by the
Fermi-type function\cite{kss}, and the $\delta R_0$-part only has been discussed,
assuming $\delta R_0=\delta R_M$.
Before interpreting the ($L-\delta R$)-correlation in detail
in the next section 3.3.3, 
the correlation between $L $ and $\delta R_a$ should be examined.

\begin{figure}[ht]
\centering{%
\includegraphics[scale=1]{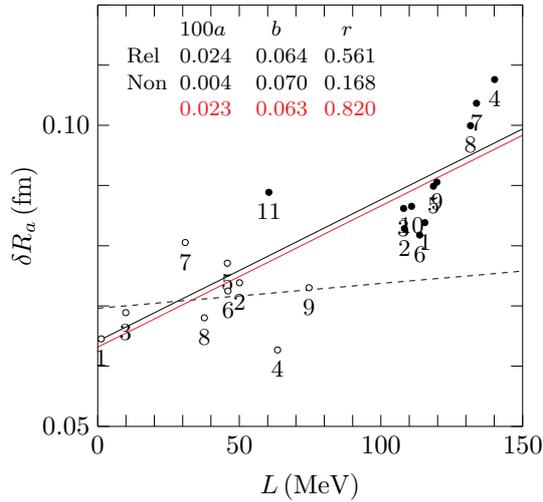}
}
\caption{
The correlation between $L$ and $\delta R_a$
in $^{208}$Pb.
Here, $L$ denotes the slope of the asymmetry energy, and $\delta R_a$ the neutron-skin thickness
coming from the diffuseness parts of the neutron and proton densities.
The closed circles show the values of $\delta R_a$ in Eq.(\ref{dra}) calculated in the RMF models
using the point neutron and proton distributions approximated by the Fermi-type functions,
while the open circles in the SMF models. Each circle is accompanied
by the number which indicates the used model specified in the text.
The black solid and dashed lines show the least square lines(LSL) for the closed and open circles,
respectively, while the red one is obtained by taking account of all the circles.
The values of the slope($a$), the intercept($b$) and the correlation coefficient($r$) of the LSL's
are written inside of the figure. For details, see the text.
}
\label{figure_Ra-L}
\end{figure}

In Fig.\ref{figure_Ra-L} is shown the ($L-\delta R_a$)-correlation,
which shows a weak correlation between $L$ and $\delta R_a$.
The values of the slopes of the LSL's are much smaller than those of the ($L-\delta R_0$)
in Fig.\ref{figure_RO-L}. The point of FSU(11) affects the slope of the LSL for the
RMF models, but 
their values of $\delta R_a$ are within $0.095\pm 0.015$ fm,
while in the SMF models,
they are within $0.070\pm 0.010$ fm. The distribution of the circles
in Fig.\ref{figure_Ra-L}
does not seem to destroy the linear correlation in Fig. \ref{figure_RO-L}.
Since almost a half of $\delta R$ is due to $\delta R_a$,
it is desirable in the future to investigate the structure of $\delta R_a$ in more detail,
for example, referring to the droplet model\cite{my,br}.

In the following, the correlation between $L$ and $\delta R$ will be discussed,
without using the Fermi-type function for the neutron and proton
distributions in the RMF and SMF models. The values of $\delta R$
are listed in Table 3 and 4.

\subsubsection{Correlation of $\delta R$ with nuclear matter parameters}\label{lsll}

\begin{figure}[ht]
\centering{%
\includegraphics[scale=1]{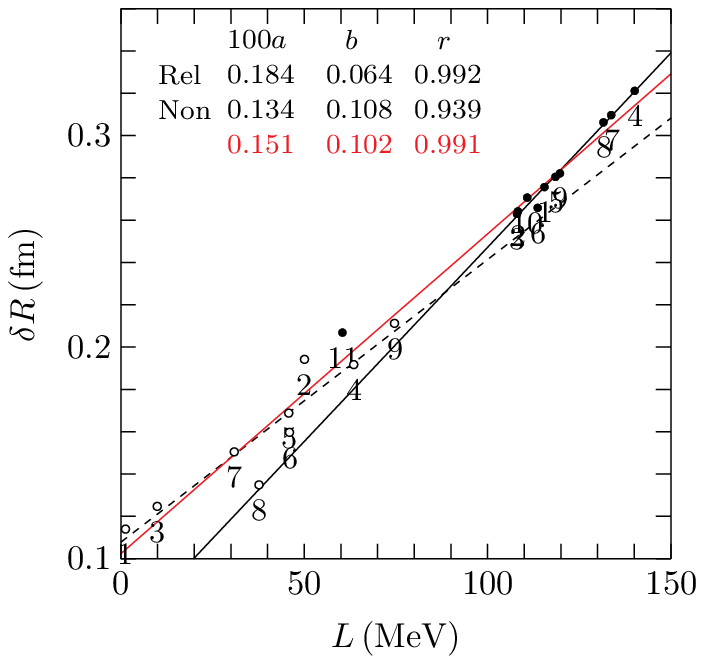}
}
\caption{
The correlation between $L$ and $\delta R$
in $^{208}$Pb.
Here, $L$ denotes the slope of the asymmetry energy, and $\delta R$ the neutron-skin thickness.
The closed circles show the values of $\delta R$ calculated in the RMF models,
while the open circles in the SMF models. Each circle is accompanied
by the number which indicates the used model specified in the text.
The black solid and dashed lines show the least square lines(LSL) for the closed
circles except for the number 11 and open circles,
respectively, while the red one is obtained by taking account of all the circles.
The values of the slope($a$), the intercept($b$) and the correlation coefficient($r$) of the LSL's
are written inside of the figure. For details, see the text.
}
\label{figure_R-L}
\end{figure}

In calculating LSL, the values given by FSU(11) are rather within the
group of the non-relativistic ones, as mentioned before. Hence, from now on,
FSU(11) will not be included in the LSA of the RMF models.
It will be included in LSA for taking all the RMF and SMF models together.

Fig.\ref{figure_R-L} shows the ($L-\delta R$)-correlation for the present choices of the MF models.
It shows almost the same correlations as in Fig.\ref{figure_RO-L} for the ($L-\delta R_0$)-correlation,
implying that the contribution from the diffuseness parts of the proton and neutron
distributions does not change so much the $(L-\delta R_0)$-relation.
The values of the LSL's in Fig.\ref{figure_R-L} are almost the same as those of the corresponding ones
in Fg.\ref{figure_RO-L}. As shown by the red line in Fig.\ref{figure_R-L},
the equation of the LSL for all the results of the RMF and SMF models
is written as $\delta R =0.151\times 10^{-2}L+0.102$ with $r=0.991$.
It should be noted that Ref.\cite{roca3} has obtained almost the same result
as $\delta R=0.147\times 10^{-2}L+0.101$ with $r=0.979$ by LSA for a much larger
number of models.

\begin{figure}[ht]
\centering{%
\includegraphics[scale=1]{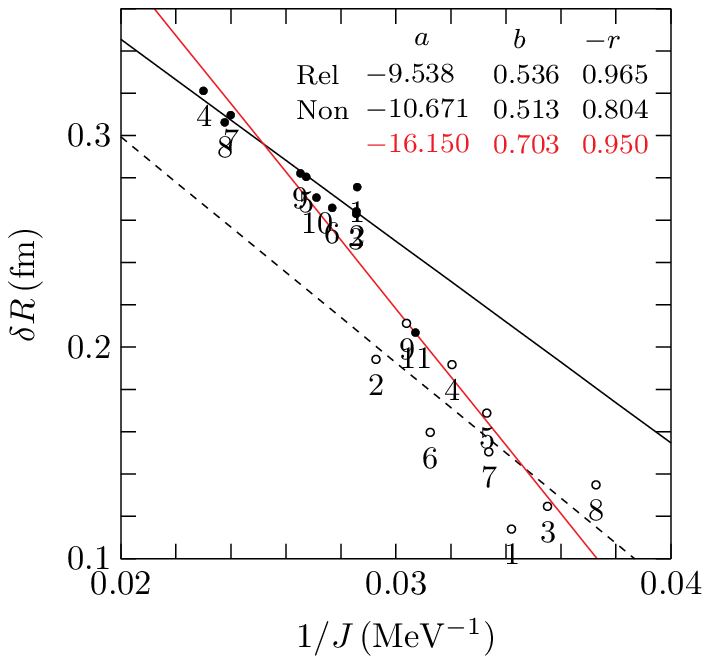}
}
\caption{
The correlation between $1/J$ and $\delta R$
in $^{208}$Pb.
Here, $J$ denotes the asymmetry energy coefficient, and $\delta R$ the neutron-skin thickness.
The closed circles show the values of $\delta R$ calculated in the RMF models,
while the open circles in the SMF models. Each circle is accompanied
by the number which indicates the used model specified in the text.
The black solid and dashed lines show the least square lines(LSL) for the closed
circles except for the number 11 and open circles,
respectively, while the red one is obtained by taking account of all the circles.
The values of the slope($a$), the intercept($b$) and the correlation coefficient($r$) of the LSL's
are written inside of the figure. For details, see the text.
}
\label{figure_R-J}
\end{figure}

The correlation between $1/J$ and $\delta R$ is shown in Fig.\ref{figure_R-J}.
It is similar to Fig.\ref{figure_RO-J}, implying that the correlation in the present case
also is dominated by the NST formula in Eq.(\ref{result}).
The values of $r$ in Fig.\ref{figure_R-J} are a little improved,
compared with those in Fig.\ref{figure_RO-J}.

\begin{figure}[ht]
\centering{%
\includegraphics[scale=1]{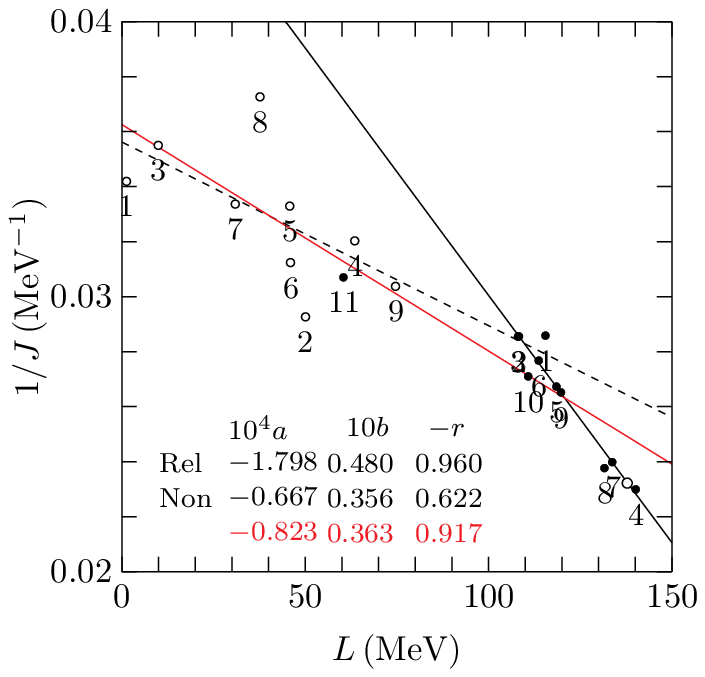}
}
\caption{
 The correlation between $L$ and $1/J$
in $^{208}$Pb.
Here, $L$ denotes the slope of the asymmetry energy, and $J$ the asymmetry energy
coefficient.
The closed circles show the values of $1/J$ in the RMF models,
while the open circles in the SMF models. Each circle is accompanied
by the number which indicates the used model specified in the text.
The black solid and dashed lines show the least square lines(LSL) for the closed
circles except for the number 11 and open circles,
respectively, while the red one is obtained by taking account of all the circles.
The values of the slope($a$), the intercept($b$) and the correlation coefficient($r$) of the LSL's
are written inside of the figure. For details, see the text.
}
\label{figure_L-J}
\end{figure}

At the end of \S 3.3.2, it has been mentioned that the correlation
between $L$ and $\delta R_0$ may be due to a spurious one on the analogy of
the correlation between $R^2_n$ and $Q^4_c$. 
The spurious correlation between $R^2_n$ and $Q^4_c$ stems from the correlation
between $R^2_n$ and $R^2_p$ by the HVH theorem in the MF models. In the same way
as in $R^2_n$ and $R^2_p$,
if the correlation between $L$ and $\delta R$ is spurious, there should be
the correlation between $L$ and $1/J$ in the MF models.
Their correlations are shown in Fig.\ref{figure_L-J},
where the closed circles and the black solid line indicate the results
of RMF models, while the open circles and the dashed line those of the SMF models.
Note that the LSL of the RMF models does not take into account FSU(11).
The red LSL is obtained by LSA for all the circles. 
Indeed, the correlation between $L$ and $1/J$ is observed in both RMF and SMF models.
The LSL's of the two frameworks are fairly different from each other,
as expected in their different $\rho$ dependences of $J$ in Eqs.(\ref{rs2})
and (\ref{ns2}).
It should be remembered for later discussions that
the slope of the LSL is extremely small.

\begin{table}
\begingroup
\renewcommand{\arraystretch}{1.2}
\hspace{2.6cm}%
{\setlength{\tabcolsep}{4pt}
\begin{tabular}{|c|c|c|c|c|c|} \hline
\multicolumn{2}{|c|}{}  &a&b&r&$\sigma$\\ \hline
$L-\delta R$   & Rel  & $0.001836$ & $0.06366$ & $0.99196$& $0.00252$ \\ \cline{2-6}
               & Non  & $0.001336$ & $0.10782$ & $0.93865$& $0.01088$ \\ \cline{2-6}
               & Rel+Non  & $0.001511$ & $0.10246$ & $0.99061$& $0.00889$ \\ \hline
$1/J-\delta R$ & Rel  & $-9.538$ & $0.536$ & $-0.965$& $0.00521$ \\ \cline{2-6}
               & Non  & $-10.671$ & $0.513$ & $-0.804$& $0.01874$\\ \cline{2-6}
               & Rel+Non  & $-16.150$ & $0.703$ & $-0.950$& $0.02029$\\ \hline
$L-1/J$       & Rel     &$-1.7984\times 10^{-4}$ &$0.04803$ & $-0.96010$ &$0.000563$ \\ \cline{2-6}
               & Non     &$-6.6723\times 10^{-5}$ &$0.03562$ & $-0.62166$ &$0.001863$  \\ \cline{2-6}
               & Rel+Non &$-8.2277\times 10^{-5}$ &$0.03625$ &$-0.91658$  &$0.001531$    \\ \hline
\end{tabular}
}
\endgroup
\caption{The values of the slope and interception of the least square lines(LSL), $\delta R=aL+b$,
 $\delta R=a(1/J)+b$ and $1/J=aL+b$, and those of their correlation coefficients ($r$) and the standard
 deviations ($\sigma$) in $^{208}$Pb.
The notation Rel stands fot the RMF models except for FSU and Non the SMF ones,
while (Rel+Non) includes FSU also. 
 The neutron-skin thickness $\delta R$ is given
 in units of fm,
 while the symmetry energy $J$ and its slope $L$ in units of MeV. The listed values are provided
so as to have appropriate units in fm and MeV in the equations of LSL. 
For details, see the text.}
\label{table_lsl}
\end{table}

In the case of the moments, the value of $R^2_n$ is correlated with the one of $R^2_p$
by the HVH theorem. Between $L$ and $J$, 
there is no reason in principle why they should be correlated with each other
in the MF framework.
Nevertheless, the impressible improvement in the sense of the value of $r$ has been observed in
Fig.\ref{figure_R-L}, compared with Fig.\ref{figure_R-J}.
In LSA, the improvement of $r$ does not necessarily have a physical meaning.
For example, such an improvement has not been observed in the LSA of the moments. 
In the ($R^2_n-R^2_c$)- and $(R^2_p-R^2_c)$-correlations, 
the standard deviation of the former in the RMF(SMF) models
is $\sigma_n=0.2170(0.2242)$ fm$^2$, while that of the latter
$\sigma_p=0.0015(0.0000)$ fm$^2$.
It may be required, however, to understood the reason of the improvement
in Fig.\ref{figure_R-L} by moving from Fig.\ref{figure_R-J} through Fig.\ref{figure_L-J}.

The equations of LSL's in the above three figures are described, respectively, as
\begin{equation}
\delta R = -a_J (1/J) + b_J\,, \quad
\delta R = a_L L+ b_L\,,\quad
1/J=-\alpha L + \beta\,.\label{lsl3}
\end{equation}
In the figures, they are written in the same notations with $a$ and $b$,
as explained in Table 5.
For example, the values of the coefficients of the red lines are given in Table 5 as
\begin{align}
a_J&=16.150\,,\quad b_J=0.703 \,,\quad (\sigma_J= 0.0203)\,,\\
a_L&=0.151\times 10^{-2}\,,\quad b_L=0.102 \,,\quad (\sigma_L= 0.00889)\,,\\
\alpha&=0.823\times 10^{-4}\,,\quad \beta=0.363\times 10^{-1}\,,\quad (\sigma_\gamma=0.00153),
\end{align}
where the values of the standard deviations are added in the parentheses, respectively.
The units of $\delta R$ are given by fm, those of $J$ and $L$ by MeV,
and others are by a proper way using fm and MeV, as those of $a_J$ by fm$\cdot$MeV.
The improvement of the $(L-\delta R)$-correlation is recognized,
comparing the value of $\sigma_L$ with that of $\sigma_J$ for the $(1/J-\delta R)$-correlation.

The calculated values of the model $i$ are described as 
\begin{equation}
\delta R_i = -a_J (1/J_i) + b_J+\delta j_i\,, \quad
\delta R_i = a_L L_i+ b_L+\delta l_i\,,\quad
1/J_i=-\alpha L_i + \beta +\delta\gamma_i\,,\label{lsli}
\end{equation}
where $\delta j_i$, $\delta l_i$ and $\delta\gamma_i$ provide the standard deviations,
\begin{equation}
n\sigma_J^2=\sum_{i=1}^n \delta j_i^2\,,\quad
 n\sigma_L^2=\sum_{i=1}^n \delta l_i^2\,,\quad
 n\sigma_\gamma^2=\sum_{i=1}^n \delta \gamma_i^2\,, \label{sd}
\end{equation}
$n$ being the number of the models.
By the definition of LSL, the following relationships hold for the mean values
of the models,
\begin{equation}
\langle \delta j_i\rangle =\langle \delta l_i\rangle
 = \langle \delta \gamma_i\rangle =0\,,\quad
\langle (1/J_i)\delta j_i\rangle =\langle L_i\delta l_i\rangle
= \langle L_i\delta \gamma_i\rangle =0\,.\quad \label{mv}
\end{equation}
The use of Eqs.(\ref{lsli}) to (\ref{mv}) yields the various relationships
between the standard deviations. One of them is written as
\begin{equation}
\sigma_L^2=\sigma_J^2+a^2_J\sigma^2_\gamma-(a_L^2-\alpha^2 a_J^2)
 (\langle L_i^2\rangle-\langle L_i\rangle^2)\label{sdl}.
\end{equation}
This equation shows that the value of $\sigma_L^2$ is increased
from $\sigma_J^2$ by $a^2_J\sigma^2_\gamma$, but is decreased by the last term
of the right-hand side. As a result, the value of $\sigma_L^2$ for the ($L-\delta R$)-relation
becomes smaller than that of $\sigma_J^2$ for the ($1/J-\delta R$)-one.
Actually, the following numbers are obtained for Eq.(\ref{sdl}) in the present case,
\begin{align}
\sigma_L^2=&(0.0088939)^2\,,\nonumber\\
\sigma_J^2+&a^2_J\sigma^2_\gamma-(a_L^2-\alpha^2 a_J^2)
 (\langle L_i^2\rangle-\langle L_i\rangle^2)\nonumber\\
&=(0.0202875)^2+(16.14975\times 0.0015306)^2\nonumber\\
 &-(0.00151142^2-8.22766^2\times 10^{-10}
 \times 16.1497^2)\times 1818.513.    \label{sdll}
\end{align}
Thus, $\sigma_L<\sigma_J$ is owing to the small value of $\alpha$
which makes the value of $(a_L^2-\alpha^2 a_J^2)$ positive
in the last term of the right-hand side of Eq.(\ref{sdl}). 
If the last equation in Eq.(\ref{lsl3}) were inserted into the first one,
then one would have $\alpha a_J=a_L$
which is for the case of $\delta j_i=\delta l_i=\delta \gamma_i=0$.

Eq.(\ref{sdl}) is also described as
\begin{equation}
\sigma_L^2=
 \sigma_J^2+a^2_J\sigma^2_\gamma
 -\frac{1}{\alpha}(a_L+\alpha a_J)\langle\delta j_i\delta\gamma_i\rangle.\label{sdj}
\end{equation}
As seen in Figs.\ref{figure_R-J} and \ref{figure_L-J}, most of the models
provide the positive values of $\delta j_i\delta\gamma_i$ for the last term
which make the value of $\sigma_L^2$ smaller.
In the LSA of the moments, $\sigma_p^2$ corresponds to the first term
of the right-hand side of Eq.(\ref{sdj}). Its value has been obtained
to be almost zero, as mentioned before.
Equivalently, the deviation from the LSL corresponding to $\delta j_i$ in the 
the last term is negligible. Hence, the value of $\sigma_n^2$ stems from
the one corresponding to the second term $a^2_J\sigma^2_\gamma$,
which is larger than $\sigma^2_p$.

The LSA for any set of the elements yields the LSL which satisfies
the equations like Eqs.(\ref{sd}) and (\ref{mv}). Hence,
without the reference formulae like Eqs.(\ref{result}), (\ref{4thm})
and (\ref{msr}), it may not be appropriate to discuss the model-dependence
of the correlations as to $L$ in detail.
A few comments, however, may be helpful for understanding the model-dependence.

Fig.\ref{figure_L-J} shows the correlation coefficient of
the RMF models is closed to 1, in contrast to that of SMF ones.
The reason of this fact may be understood as follows.
Eq.(\ref{L2}) is described as
\begin{equation}
L=3J+3\rho_o^2\frac{\partial}{\partial \rho}\left(\frac{J(\rho)}{\rho}\right)|_{\rho=\rho_0, \rho_3=0}
\label{L3}.
\end{equation}
For the relativistic framework excluding FSU(11),
Table 3 gives the average values of $L_i$ and $J_i$
satisfying
\begin{equation}
\langle L_i\rangle\approx 3\langle J_i\rangle\,\,\,,
 \quad (\langle L_i\rangle\approx 119.999\,,
 3\langle J_i\rangle\approx 114.105),
\end{equation}
which implies the second term of Eq.(\ref{L3}) is small. Then, in writing 
$L_i=3J_i+\Delta_{L_i}$,\,($\Delta_i\ll3J_i$),  $1/J_i$ is expanded as 
\begin{equation}
1/J_i\approx -\frac{3}{\langle L_i \rangle^2}L_i+\frac{3}{\langle L_i \rangle}
 \left(2+\frac{\langle\Delta_i\rangle}{\langle L_i \rangle}\right)
\end{equation}
Table 3 provides the values for the right-hand side of the above equation,
\begin{equation}
\frac{3}{\langle L_i \rangle^2}=2.083\times 10^{-4}\,,\quad
\frac{6}{\langle L_i \rangle}=5.000\times 10^{-2},
\end{equation}
which are comparable with the values of $\alpha=1.798\times 10^{-4}$
and $\beta=4.803\times 10^{-2}$ for the LSL of the ($L-1/J$) correlation in Table 5.
As to Eq.(\ref{ns2}) for the SMF models, a brief discussion similar to the above is not possible.

\begin{figure}[h]
\begin{minipage}{7.7cm}
\includegraphics[bb=0 0 211 161]{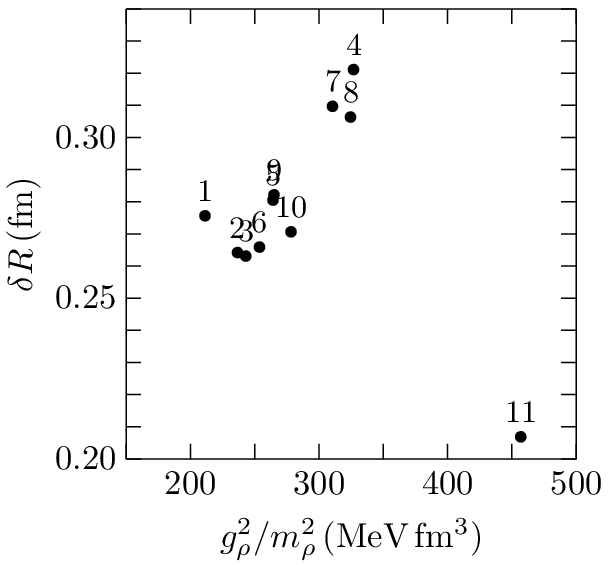}%
 \caption{The correlation between the neutron-skin thickness($\delta R$)
and the strength of the $\rho$-meson exchange interaction($g^2_\rho/m^2_\rho$)
in the RMF models.
 Each circle is accompanied
by the number which indicates the used model specified in the text.
 }
\label{rmf}
\end{minipage}\hspace{0.3cm}%
\begin{minipage}{7.7cm}
\includegraphics[bb=0 0 211 190]{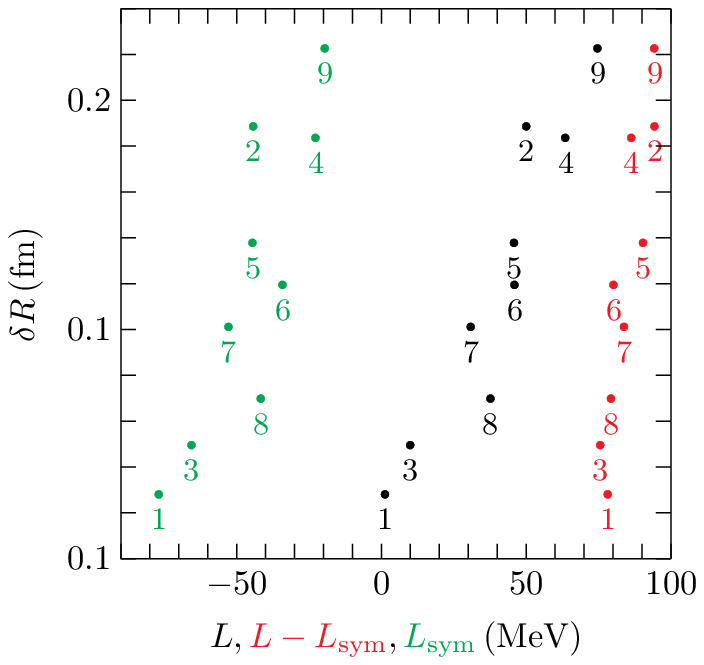}
  \caption{The correlation between the neutron-skin thickness($\delta R$)
and the slope of the asymmetry energy($L=L_{\rm non}$) in the SMF models.
For the definition of the $L_{\rm sym}$, see the text.
The colors of the circles correspond to those of the quantities indicated on the bottom.
Each circle is accompanied by the number which stands for the used model specified in the text.
}
\label{smf}
\end{minipage}
\end{figure}

A common feature of the RMF and SMF models in Fig.\ref{figure_R-L} 
is that $\delta R$ is an increasing function of $L$. The structures of $L$ in the two models, however,
are different from each other, as follows.

On the one hand, the function of $L$
in the RMF models($L=L_{\rm rel}$) is given by Eq.(\ref{rs2}) for $\lambda=0$ as 
\begin{equation}
L_{\rm rel}=L_k+\frac{3}{2}\frac{g^2_\rho}{m^2_\rho}\rho_0,\label{rmfl}
\end{equation}
where the first term of the right-hand side comes from the first one of Eq.(\ref{rs2}),
and is almost independent of the increasing correlation of $\delta R$ with $L$.
Fig.\ref{rmf} shows the correlation of the $\delta R$ with the second term of the above
equation. It is seen that the increasing
of $\delta R$, except for that of FSU(11), is dominated by the second term
from the $\rho$-exchange interaction in Eq.(\ref{rmfl}).

On the other hand, the part of $L=L_{\rm non}$ in the SMF models,
which is responsible for the increasing of $\delta R$, is found by writing Eq.(\ref{ns2}) as
\begin{equation}
L_{\rm non}=(L_{\rm non}-L_{\rm sym})+L_{\rm sym},\label{smfl}
\end{equation}
where $L_{\rm sym}$ is provided by the curvature of the asymmetry energy($K_{\rm sym}$) as,
\begin{equation}
L_{\rm sym}=\frac{K_{\rm sym}}{3(\alpha+1)},\quad
K_{\rm sym}=3\rho^2_0\frac{\partial}{\partial\rho}\left(\frac{L}{\rho}\right)|_{\rho=\rho_0, \rho_3=0}.
\end{equation}
Fig.\ref{smf} shows the contribution from each component
of Eq.(\ref{smfl}) to $\delta R$. It is seen that the increasing correlation
of $\delta R$ with $L$ is due largely to $L_{\rm sym}$ given by the green points.
In the SMF model, the values of $L_{\rm sym}$ are negative as in Fig.\ref{smf},
while Ref.\cite{roca2} has shown that there are several RMF models as NL3(5)
and NL-SH(6) with the positive values of $K_{\rm sym}$.
Thus, there seems to be
no clear
constraint on the structure of $L$ in the MF models.

From the discussions in \S 3.3, it is concluded that the linear relationship
of $L$ with $\delta R$ is spurious through the correlation of $L$ with $J$
which is one of the main components of the NST formula in Eq.(\ref{result}).
In the case of LSA on the moments, the value of $R^2_n$ determined by the
spurious correlations with $R^2_c$ and $Q^4_c$ is confirmed by the correlation
with $Q^4_{cn}$ whose value is obtained through experiment.
For $L$, there is not such an observable by which
the value of $L$ obtained from the spurious correlation is confirmed.  
If the experimental value of $\delta R$ is determined, then the value of 
the symmetry energy coefficient $J(\rho_0)$ will be fixed 
by LAS consistently in the model-framework.
Since the $\rho$-dependence of $J(\rho)$ is not known,
the value of $L$ is given by the assumed $J(\rho)$ model-dependently,
according to Eq.(\ref{L2}) with the equation like Eq.(\ref{rs2}), (\ref{ns2}),
and others. The improvement of the $(L-\delta R)$-correlation in the present LSA,
compared with the ($1/J-\delta R$)-one, is due to
the ($1/J-L$)-correlation which is responsible for the spurious
correlation.

\section{Summary}

The neutron-skin thickness(NST) of the asymmetric semi-infinite nuclear
matter(SINM) has been shown in an analytic way to be expressed in terms of macroscopic
quantities of nuclear matter. They are the asymmetry energy coefficient($J$),
the slope($L$) of the asymmetry energy($J(\rho)$), and the incompressibility
coefficient($K$), in addition to the Fermi momentum($k_{\rm F}$),
the Coulomb energy($V_c$) and the asymmetry factor(I=(N-Z)/A).
Here, the notation $\rho$ indicates the nucleon density in the nuclear matter.
The NST formula, which is given by Eq.(\ref{result}), is derived according to
the Hugenholtz-Van Hove(HVH) theorem in the mean-field(MF) approximation, 
and is independent of details of the interaction parameters in the MF
phenomenological models.

The NST formula is used as a guide to investigate the NST of $^{208}$Pb
defined by $\delta R=R_n-R_p$ where $R_n$ and $R_p$ denote the root-mean-square
radius of the point neutron and proton distributions, respectively.
The NST of finite nuclei is separated into two parts
as $\delta R=\delta R_0+\delta R_a$ by using the Fermi-type distribution
of neutrons and protons. The term $\delta R_0$ corresponds to the NST of SINM,
and $\delta R_a$ is due to the diffuseness part of the Fermi-type function.
The value of $\delta R_0$ is dominated by $V_c/J$,
as in conventional understanding in the literature\cite{bm}, while 
the contributions of $L$ and $K$ to $\delta R_0$ are less than $10\%$.

In the least squares analysis(LSA) of the calculated values of $\delta R$
in the relativistic(RMF) and non-relativistic (SMF) MF models for $^{208}$Pb,
the well defined least square line(LSL) is obtained in the $(1/J-\delta R)$-plane,
employing $20$ models cited in Ref.\cite{ks2}.
The LSL between $L$ and $\delta R$ is also observed in the $(L-\delta R)$-plane,
but its correlation is a kind of spurious ones, through the correlation
of $L$ with $1/J$ which is one of the main components of $\delta R_0$
in the NST formula.
About half of $\delta R$ is due to $\delta R_a$, but the ($L-\delta R$)-correlation
is insensitive to $\delta R_a$.

In the case of $R^2_n$, the spurious correlation with 
the charge radius($R^2_c$) appears 
owing to the constraint on $\delta R$ by the HVH theorem, but the value of $R^2_n$
is confirmed by the value of the fourth moment of the neutron charge density($Q^4_{cn}$)
estimated from experiment.
For $L$, there is not such a  constraint on the ($1/J-L$)-relationship,
and no observable related to $L$ like $Q^4_{cn}$ for $R^2_n$.
As a result, the value of $L$ remains to be a free parameter in the MF models.

A few comments may be useful for the present paper.
First, in the LSA, it may be necessary to have an established reference formula,
like Eq.(\ref{result}) for $\delta R$ and Eq.(\ref{4thm}) for $R^2_n$.
Otherwise, the meaning of the LSL obtained numerically is not clear,
and it is not easy to distinguish
spurious correlations from others.

Second, the value of the asymmetry energy coefficient $J$ is determined
at the saturation density $\rho_0$, as $J(\rho_0)$, and hence, does not provide
any information on
the density-dependence of $J(\rho)$. As a result, the relationship between the
slope $L$ and $J(\rho)$ is model-dependent.
This is the same as the fact that
there is no way to determine the value of $K$ from the energy density at $\rho_0$.
In the determination of the value of $K$, additional experimental data are required,
such as the excitation energy of the breathing-mode oscillation\cite{bla}.
In the same way, in addition to $\delta R$, other observables may be required for
estimating the value of $L$.

Third, even if the value of $\delta R$ is fixed experimentally,
the present LSA is not a kind of methods to determine the
experimental value of $J(\rho_0)$,
but provides the value of $J(\rho_0)$ allowed in each model-framework
used in the LSA\cite{kss}.
There is no reason why the RMF and SMF should be analyzed together.
This fact has shown explicitly in Ref.\cite{kss}, where
the values of $R^2_p$ in the RMF and SMF models are determined by the LSA,
using the experimental value of the charge radius from electron scattering.

In the present paper, each framework allows different value of $J$.
Ref.\cite{kss} has provided $\delta R = 0.275(0.070)$ fm for the RMF models, and
$0.162(0.068)$ fm for the SMF ones. For these values,
the present LSA for the ($1/J-\delta R$)-correlations
yields  $J=39.9(11.5)$ and $32.4(8.0)$ MeV, respectively.
If the LSL obtained for the ($L$-$\delta R$)correlation is used,
it provides  $L=115.1(39.5)$ and $40.6(59.0)$ MeV
for the two frameworks, respectively, including the standard deviation
of the LSL lines.

It should be noted that the recent paper from JLab\cite{jlab2}
gives $\delta R=0.283(0.071)$ fm, using a similar LSA on the data of
parity-violating electron scattering(PVES) experiment, while Ref.\cite{jlab3}
$0.19(0.02)$ fm, using the same JLab data, but with the different
model-dependent analysis. The former is almost equal to the value 
of the relativistic models in Ref.\cite{kss},
while the latter close to that of the non-relativistic ones. 
Unfortunately, however, it is not possible to make a clear discussion
on their PVES results, comparing  with those in Ref.\cite{kss},
since the former analyses have taken into account the
relativistic and non-relativistic models together in LSA, using the PVES data
obtained by the phase shift analysis which does not distinguish
the different moments\cite{ks1}.

Finally, the relationship between Ref.\cite{ks2} and the present paper
should be mentioned.
Ref.\cite{kss} has shown that the value of $\delta R$ is larger by about $0.1$ fm
in the RMF models than in the SMF ones.
This difference has been understood in Ref.\cite{ks2} by the HVH equation,
\begin{equation}
\langle m^\ast_{\tau}\rangle\langle V_\tau\rangle
\approx a_\tau+b_\tau\langle m^\ast_{\tau}\rangle,\nonumber\label{hvheq}
\end{equation}
derived from the HVH theorem. Here,
$\langle m^\ast_{\tau}\rangle$ stands for the average nucleon effective mass
in units of $M$,
and $\langle V_\tau\rangle$ the average nuclear potential in each framework. 
The subscript stands for $\tau =n$ for neutrons, and $\tau=p$ for protons.
The values of the constants, $a_\tau$ and $b_\tau$, 
depend on the average values of $\rho_\tau$($\langle \rho_{\tau}\rangle$),
the binding energy per nucleon $\eb$ and Coulomb energy $V_c$
of the corresponding asymmetric nuclear matter with $N$ and $Z$.
Since the values of $\eb$ and $V_c$ are almost the same in the RMF
and SMF models, the difference between the two frameworks in the right-hand side
of the HVH equation is attributed to the difference between the values
of $\langle \rho_\tau\rangle$ and $\langle m^\ast_\tau\rangle$.
Then, the difference between $\delta R$'s is shown to be approximately given
by the ratio of $(\langle m^\ast_n\rangle\langle V_n\rangle)^{1/4}$ in the two frameworks.
Indeed, this ratio calculated by the RMF and SMF models explains well
the $0.1$ fm difference\cite{ks2}.
In contrast to the above HVH equation, the present paper has derived $\delta R$
directly from $\rho_\tau$ in terms of $k_{\rm F}$, $I$, $J$, $L$ and $K$.
As a result, the $0.1$ fm difference is
attributed mainly to the difference between the $J$-values,
instead of  $\langle m^\ast_n\rangle $ and $\langle V_n\rangle$.

\section*{Acknowledgments}

The author would like to thank Professor H. Kurasawa for valuable discussions.
Without his substantial cooperation, this paper would not have been completed.
The continuous support by Professor T. Suda was also indispensable to this work.

\end{document}